\documentclass[12pt]{iopart}

\usepackage{graphicx}
\usepackage[T1]{fontenc} 

\expandafter\let\csname equation*\endcsname\relax
\expandafter\let\csname endequation*\endcsname\relax

\usepackage{amsmath,amssymb} 

\usepackage{braket} 
\usepackage{physics} 

\usepackage{cite}
\usepackage{color}

\begin{document}

\title{Recent advances in hole-spin qubits}

\author{Yinan Fang$^1$, Pericles Philippopoulos$^2$, Dimitrie Culcer$^{3,4}$, W. A. Coish$^5$, and Stefano Chesi$^{6,7}$}

\address{$^1$School of Physics and Astronomy, Yunnan University, Kunming 650091, China}
\address{$^2$Nanoacademic Technologies Inc., Montr\'eal QC, Canada}
\address{$^3$School of Physics, University of New South Wales, Sydney NSW 2052, Australia}
\address{$^4$Australian Research Council Centre of Excellence in Future Low-Energy Electronics Technologies, The University of New South Wales, Sydney NSW 2052, Australia}
\address{$^5$Department of Physics, McGill University, Montr\'{e}al QC, Canada}
\address{$^6$Beijing Computational Science Research Center, Beijing 100193, China}
\address{$^7$Department of Physics, Beijing Normal University, Beijing 100875, China}

\ead{stefano.chesi@csrc.ac.cn}

\begin{abstract}
In recent years, hole-spin qubits based on semiconductor quantum dots have advanced at a rapid pace. We first review the main potential advantages of these hole-spin qubits with respect to their electron-spin counterparts, and give a general theoretical framework describing them. The basic features of spin-orbit coupling and hyperfine interaction in the valence band are discussed, together with consequences on coherence and spin manipulation. In the second part of the article we provide a survey of experimental realizations, which spans a relatively broad spectrum of devices based on GaAs, Si, or Si/Ge heterostructures.  We conclude with a brief outlook. 
\end{abstract}

%
%
%
%
%

Nearly 25 years have passed since the idea of implementing quantum information processing with semiconductor quantum dots was put forward~\cite{LDV98}. The spin of confined electrons (or holes) is a natural two-level system, relatively well shielded from charge fluctuations and which, as it has been now established, can be coherently manipulated with high accuracy and speed. Research in the field has been constantly progressing over the years, exploring at the same time the rich physics of these semiconductor devices, but is still restricted  to the few-qubits regime. At the moment it is not yet obvious which platform offers better prospects to realize intermediate- or large-scale devices, and recently hole-spin devices have gained prominence.

Recent rapid progress in controlling hole spins has brought breath of fresh air to the field. However, the potential advantages (and disadvantages) of these qubits over their electron-spin counterparts  should be carefully weighted. To get some perspective, it is interesting to discuss how this family of spin qubits fits the main line of progress, dominated by electron-spin devices. Within about 10 years from the 1998 theoretical proposal, all the main ingredients required to realize quantum information processing, such as spin readout, single-spin manipulation, and controllable spin-spin exchange interactions, were firmly established. An account of early progress can be found in the 2007 review Ref.~\cite{Hanson07}. In this initial phase, the most successful qubits were planar (lateral) quantum dots in AlGaAs/GaAs heterostructures. However, these systems are susceptible to strong dephasing caused by nuclear spins. This dephasing is unavoidable in III-V semiconductors since every isotope of Ga and As has a finite nuclear spins. To remedy this situation, Si electron-spin devices have gradually emerged as a better alternative. Coherent singlet-triplet oscillations in a Si/SiGe double quantum dot were demonstrated in 2012 \cite{Maune2012} and a number of landmark results have been obtained so far, such as high-fidelity ($>99.9\%$) single-qubit manipulation \cite{Tarucha18,Dzurak19}, strong coupling of spin qubits to a superconducting resonator \cite{Samkharadze18,Mi18,Borjans19,Harvey22} and, very recently, two-qubit quantum processors with gate fidelities surpassing the fault-tolerance threshold~\cite{Xue22,Noiri22,Mills22} and universal operation of a six-qubit array~\cite{Vandersypen22}.

Hole-spin qubits fit naturally into the historical path, as they can overcome the influence of nuclear spins in the same way as electron-spin devices based on Si: For hole-spin qubits in Si and Ge, the low natural abundance of spinful nuclei can be further reduced by isotopic purification. Furthermore, even in III-V semiconductors, holes are much less affected by nuclear spins than conduction-band electrons. The Fermi contact interaction vanishes, leading to an anisotropic hyperfine interaction caused by dipolar and orbital terms, weaker by about an order of magnitude. In addition to having a weaker influence of nuclear spins, two attractive features of holes should facilitate scalability. The first is the absence of a valley degree of freedom. The valley splitting of electrons in Si quantum dots depends on atomistic features of the device, which are difficult to control uniformly over large quantum-dot arrays. The second and most important feature is the strong spin-orbit interaction of holes, allowing in-situ electric control of spin couplings and effective spin manipulation through a local electric drive. In contrast, the spin-orbit coupling of conduction electrons in Si is weak, and single-spin manipulation (as well as coupling of the spin to superconducting resonators) requires large auxiliary elements, such as micromagnets \cite{Tokura06,Pioro08}.

Strong spin-orbit coupling is, of course, a double-edged sword, as it enhances the influence of charge noise on the spin degrees of freedom. Despite this general difficulty, and the late start of hole devices in terms of fabrication, in recent years hole-spin qubits have achieved remarkable results. In many aspects concerning single-spin manipulation \cite{Veldhorst20,Froning2021,Guo22a, Veldhorst21_arXiv}, scalability~\cite{Veldhorst21,Veldhorst22}, as well as coupling to superconducting circuits \cite{Franceschi22a}, hole-spin qubits can approach and sometimes already surpass the performance of electron-spin qubits. 

To describe recent progress in hole-spin qubits, we have structured the article in two main parts. Section~\ref{sec:theory} gives a theoretical perspective, introducing the general framework for hole-spin qubits and the main interactions affecting them. Section~\ref{sec:experiments} is a survey of various experimental platforms, highlighting some of the most interesting achievements. While the present review does not aim to be complete, it should offer an informative and sufficiently broad view of hole-spin qubits. Recently, two more specialized reviews have appeared: Ref.~\cite{Studenikin21review}, focusing on hole spins in GaAs lateral quantum dots, and Ref.~\cite{Scappucci20}, reviewing hole-based Ge devices. For a more complete perspective on spin qubits (including electron-spin systems), one could consult Ref.~\cite{Burkard21_review}.

\section{General description of hole-spin qubits}\label{sec:theory}

Quantum dots for use in information processing are typically based on either III-V compound semiconductors such as GaAs, which have zincblende lattices, or on group-IV semiconductors such as Si and Ge, with diamond lattices. While the conduction band is formed by $s$-like atomic wave functions, resulting in spin-$1/2$ quasiparticles, valence-band states in diamond and zincblende semiconductors are predominantly formed by atomic $p$-orbitals. In the atomic limit, combining the $L=1$ orbital angular momentum with the electron spin-$1/2$ we obtain two possibilities for the total angular momentum $J$, henceforth also referred to as (effective) spin: $J=3/2$ or $J=1/2$. At zero crystal momentum ${\bf k}$ these states are separated by a gap and the band originating from the $J=1/2$ manifold, which is always lower in energy, is known as the \textit{split-off} band. More precisely, since rotational symmetry is broken in the crystal, the valence-band states at ${\bf k}=0$ are classified by the $\Gamma_8$ and $\Gamma_7$ representations of the tetrahedral double group $T_d$ \cite{Kostner63,Dresselhaus08}. Furthermore, the crystalline potential leads to the hybridization of $p$ orbitals with $d$ states \cite{Chadi76,Boguslawski94,Bassani98,Zunger03,Diaz06}, which can significantly affect the form of the hyperfine interaction \cite{Chekhovich13b,Philippopoulos20} (see Section~\ref{sec:hyperfine}). Despite these caveats, the fourfold degeneracy of atomic $J=3/2$ states (corresponding to $\Gamma_8$) is preserved. The existence of the split-off band makes a difference for numerical values, especially in Si, but properties of holes are usually well understood considering the states at the top of the valence band. Therefore, the following Luttinger Hamiltonian\cite{Luttinger56, winkler2003spin} is often taken as a fundamental starting point for the description of holes:  
\begin{eqnarray}\label{HLut}
 H_{L}({\bf k}) = && -\displaystyle \bigg(\gamma_1 + \frac{5}{2}\, \gamma_2\bigg) \, \frac{\hbar^2 {\bf k}^2}{2m_0} 
+ \gamma_2 \frac{\hbar^2}{m_0} \, (k_x^2 J_x^2 + k_y^2 J_y^2 + k_z^2 J_z^2) \nonumber \\
&& +\gamma_3 \frac{\hbar^2}{2m_0} \, (\{ k_x, k_y \} \{ J_x, J_y \} + {\rm c.p.}),
\end{eqnarray}
where $\{J_i\}$ are spin-$3/2$ matrices, $\gamma_1$, $\gamma_2$ and $\gamma_3$ are the Luttinger parameters, $m_0$ is the bare electron mass,  $\{ A,B\}=(AB+BA)/2$, and c.p. are terms obtained from cyclic permutations of $x,y,z$.
In contrast with conduction electrons, where perturbative spin-orbit interactions arise from the valence band and are suppressed by the large band gap of the semiconductor, the Hamiltonian describing the valence-band holes [Eq.~(\ref{HLut})] features a strong spin-orbit coupling between ${\bf k}$ and ${\bf J}$.
This coupling results in a splitting of the fourfold degeneracy at ${\bf k}=0$ into two doubly degenerate bands at finite momentum. The higher (lower) valence band states are denoted as heavy (light) holes, with angular momentum $J_z=\pm 3/2$ ($\pm 1/2 $) in the limit ${\bf k}\to 0$. Since $\gamma_{2,3}$ are generally of the same order as $\gamma_1$, the spin-orbit splitting between heavy-hole and light-hole bands is comparable to the total kinetic energy.

\emph{Envelope function approximation.} In the presence of smooth external potentials and magnetic fields, the microscopic Bloch wave at ${\bf k}=0$ is modulated by a slowly-varying envelope that satisfies a multi-component Shr\"{o}dinger equation \cite{Kohn1955EMA, winkler2003spin}. The usual prescription for \emph{electronic} states is to replace $\hbar{\bf k}\to {\boldsymbol{\pi}} = -i\hbar \boldsymbol{\nabla} + e{\bf A}({\bf r})$ ($-e<0$ is the electron charge) in $H_L$, where ${\bf A}({\bf r})$ is the vector potential associated with the external magnetic field ${\bf B}$. As usual, the components of the canonical momentum ${\boldsymbol{\pi}}$ do not commute, in general, thus in the second line of Eq.~(\ref{HLut}) we have specified the appropriate symmetrized products of ${\bf k}$ (which become unnecessary when ${\bf B} =0$). Some care should be taken in writing the Hamiltonian for the state of a missing electron. First, the electronic energy is subtracted from a filled valence band, which can be simply accounted for by a global change of sign in the Hamiltonian. This leads to the following Hamiltonian describing holes within the envelope-function approximation:
\begin{equation}\label{Henv_funct}
H_{env}=  -H_L(\boldsymbol{\pi}/\hbar)+eV({\bf r}) 
+2 \kappa \mu_{\rm B} {\bf J}\cdot {\bf B}  +2 q\mu_{\rm B} \boldsymbol{\mathcal{J}} \cdot {\bf B},
\end{equation}
where $e V({\bf r})$ is the 3D confining potential and the last two terms describe the Zeeman interaction \cite{Luttinger56, winkler2003spin}, with $\boldsymbol{\mathcal{J}}=\left(J_x^3,J_y^3,J_z^3\right)$ and $\mu_{\rm B}$ the Bohr magneton. While Eq.~(\ref{Henv_funct}) is a legitimate choice, giving the correct energy of the hole states, it still refers to the electronic wavefunction and this should be taken into account when computing certain observables. Recall, in particular, that a hole with quantum numbers $({\bf k},J_z)$ corresponds to a missing $(-{\bf k},-J_z)$ electronic state. To obtain directly the hole wavefunction, whose angular momentum corresponds to the physical state, one can consider an alternative version of $H_{env}$, by applying time-inversion  to it. Since $-i\hbar \boldsymbol{\nabla} \to +i\hbar \boldsymbol{\nabla}$ upon time inversion, the transformed Hamiltonian recovers the  Peierls substitution appropriate for a positively charged particle, i.e., $\boldsymbol{\pi} \to -i \hbar\boldsymbol{\nabla} - e {\bf A}({\bf r})$ in Eq.~(\ref{Henv_funct}). Furthermore, the second and third terms of Eq.~(\ref{Henv_funct}) change sign, since ${\bf J} \to -{\bf J}$.

\emph{Inversion-asymmetry and strain.} In III-V zincblende materials, for which there is no center of inversion symmetry, it is sometimes important to supplement Eq.~(\ref{HLut}) with additional terms. These corrections are known as Dresselhaus spin-orbit couplings and include the $k$-linear term:
\begin{equation}\label{Ck_term}
H_1 = \frac{2}{\sqrt{3}} C_k \left[k_x \{J_x, J_y^2-J_z^2 \}+{\rm c.p.}\right],
\end{equation}
and a more convoluted $k$-cubic contribution \cite{winkler2003spin}: 
\begin{align}\label{k3_term}
H_3= & b_{41} \left[\{k_x, k_y^2-k_z^2 \} J_x +{\rm c.p.}\right]+ b_{42}  \left[\{k_x, k_y^2-k_z^2 \} J^3_x +{\rm c.p.}\right] \nonumber \\
&+ b_{51}  \left[\{k_x, k_y^2+k_z^2 \} \{J_x, J_y^2-J_z^2 \} +{\rm c.p.}\right]+b_{52}  \left[k_x^3 \{J_x, J_y^2-J_z^2 \} +{\rm c.p.}\right].
\end{align}
In particular, the $k$-cubic terms were shown early on to give a significant spin-orbit splitting in symmetric GaAs quantum wells \cite{Sherman88}. A recent theoretical analysis \cite{Coish20} has assessed the relative importance of $H_1$ over $H_3$ in GaAs 2DHGs. Furthermore, a different type of coupling induced by inversion-asymmetry, but also depending on the external potential $V({\bf r})$, has been discussed~\cite{Coish20}:
\begin{equation}\label{Hdipole}
H_{E}= \frac{e a_B \chi}{\sqrt{3}}[E_x \{ J_y, J_z\} + {\rm c.p.}],
\end{equation} 
where ${\bf E} = - \boldsymbol{\nabla}V$ is the local electric field, $a_B$ is the Bohr radius, and $\chi$ is a material parameter. Equation~(\ref{Hdipole}), which involves the gradient of the potential and includes interband matrix elements between heavy- and light-hole states,  goes beyond the usual envelope-function approximation. Such pseudospin-electric coupling physically originates from the interaction between the electric field and the dipole operator, which takes the general form of Eq.~(\ref{Hdipole}) when expressed in the subspace of valence-band states. A first-principles evaluation gives a relatively large value $\chi \simeq 0.2$ in GaAs \cite{Coish20}. 

For Si and Ge, the above Dresselhaus terms are absent in the bulk. However, it is important to account for finite strain when modelling, e.g., high-mobility SiGe/Ge/SiGe quantum wells \cite{Culcer21,Veldhorst22theory} as well as core-shell nanowires \cite{Kloeffel_PRB2011,KloeffelPRB2018}. In general, the effect of strain is described by the well-known Bir-Pikus Hamiltonian $H_{\varepsilon}=-(a_v+\frac54 b_v) (\varepsilon_{xx}+\varepsilon_{yy}+\varepsilon_{zz})+b_v (J_x^2 \varepsilon_{xx} + {\rm c.p.})+\frac{2}{\sqrt{3}}d_v (\{J_x,J_y \}\varepsilon_{xy} +{\rm c.p.})$, where $\left\{ \varepsilon_{ij}\right\}$ are the strain tensor elements and $a_v,b_v,d_v$ are material-specific deformation potentials \cite{BirPikus74}. Assuming that only hydrostatic and uniaxial strain are present, i.e., $\varepsilon_{xy}=\varepsilon_{yz}=\varepsilon_{zx}=0$ and $\varepsilon_{xx}=\varepsilon_{yy}$, the effect of $H_\varepsilon$ is to induce different offsets for the heavy-hole ($J_z=\pm 3/2$) and light-hole ($J_z=\pm1/2$) states \cite{Kloeffel_PRB2011,Culcer21,Veldhorst22theory}.

\subsection{Spin-orbit interaction of confined holes }\label{sec:SOC_general}

The description of confined hole systems can be based directly on the general four-band formalism just described, eventually supplemented by the split-off band. However, many of the hole-spin qubits of interest are fabricated from low-dimensional systems such as quantum wells or nanowires, where additional gates realize the zero-dimensional confinement. It is therefore useful to describe the effective Hamiltonians governing such 2D or 1D systems. Because such effective models are relatively simple, they are often taken as starting points to characterize various important effects, such as spin relaxation or electrical manipulation of spin qubits.


\emph{Quantum wells.} For quasi-2D holes the potential in Eq.~(\ref{Henv_funct}) may be taken as $V(z)$ if the growth direction is along a main crystal axis. Then, for ${\bf B}=0$ and zero in-plane wavevector, ${\bf{k}}_{\parallel} = (k_x, k_y) = (0, 0)$, Eq.~(\ref{Henv_funct}) simplifies considerably, giving two decoupled Schr\"odinger equations in 1D. The two sectors correspond to the $J_z=\pm 3/2$ and $J_z=\pm 1/2$ subspaces, characterized by different effective masses $m_0/(\gamma_1\pm 2\gamma_2)$. The lowest-energy subband is formed by heavy-holes ($J_z=\pm 3/2$) and the energy gap to  the lowest light-hole subband can be estimated as $\Delta_{\rm LH}\simeq 2\gamma_2(\pi\hbar/L)^2/m_0$, assuming an infinite potential well of width $L$.  For more realistic scenarios, the value of $\Delta_{\rm LH}$ depends of the detailed shape of $V(z)$ and, as discussed already, is modified by uniaxial strain. In general, the  four-dimensional degeneracy of the Luttinger Hamiltonian has been broken by the confining potential, generating naturally a two-level pseudospin. For the low-energy states, we can identify the effective spin up/down eigenstates with the $J_z=\pm 3/2$ eigenvalues, i.e., $\ket{\Uparrow} = \ket{+3/2}$ and $\ket{\Downarrow} = \ket{-3/2}$. 

Despite the simple decoupling described above (which, we stress, is exact only at zero in-plane momentum), the light-hole/heavy-hole mixing terms still play a crucial role in the effective 2D Hamiltonian. As seen from Eq.~(\ref{HLut}), finite values of $k_{x,y}$ will induce matrix elements between the two subspaces, through the action of the $J_{x,y}$ operators. It is this coupling between light- and heavy-hole subbands what generates the desired effective spin-orbit interactions, to be harnessed for quantum computing. Typically, the light-hole/heavy-hole splitting $\Delta_{\rm LH}$ is much larger than energy scales associated with the in-plane motion (such as the 2DHG Fermi energy or, for spin qubits, the quantization energy of lateral quantum dots). Therefore, relevant values of $k_{x,y}$ are sufficiently small to only induce a weak mixing between heavy-holes and light-holes. In this limit, the pseudospin degree of freedom of the lowest-energy subband can still be associated with the $J_z= \pm 3/2$ quantum number, and the effective spin-orbit interactions can be conveniently derived by performing a perturbative elimination of higher-energy bands, e.g., with a Schrieffer-Wolff transformation. We list below some of the most relevant terms obtained through this procedure: 
\begin{align}\label{Hsoc}
H^{\rm 2D}_{SOI}=&\beta_D(\pi_+ \sigma_- +\pi_-\sigma_+) + \beta_{3} (\pi_+\pi_-\pi_+\sigma_- +\pi_-\pi_+\pi_-\sigma_+)\nonumber \\
& +i\alpha_R (\pi_+^3\sigma_- - \pi_-^3\sigma_+)+ \gamma (B_+ \pi_+^2 \sigma_-+B_- \pi_-^2 \sigma_+),
\end{align}
where $v_\pm = v_x \pm i v_y, {\boldsymbol{v}} \in \{{\boldsymbol{\sigma}}, {\bf{B}},{\boldsymbol{\pi}}\}$. The first line of Eq.~(\ref{Hsoc}) lists the linear and cubic Dresselhaus terms, which originate from the lack of inversion symmetry of the crystalline potential, i.e., they derive from $H_1$, $H_3$, and $H_E$ \cite{Coish20}. The spin-orbit interactions listed in the second line appear also in crystals with a center of inversion (such as Si and Ge), and the form given here is obtained from the spherical approximation of the Luttinger Hamiltonian (i.e., assuming $\gamma_2 = \gamma_3$). Specifically, the $\alpha_R$ term is the Rashba spin-orbit coupling of heavy-holes, caused by an asymmetric confinement potential [$V(z) \neq V(-z)$]. In contrast to electrons, this type of Rashba spin-orbit coupling is cubic in momentum, rather than linear \cite{Winkler00rashbaholes,Winkler02}. The last term of Eq.~(\ref{Hsoc}), which is proportional to the in-plane components of the magnetic field ${\bf B}$, was found to be relevant when modeling transport of holes in quantum point-contacts \cite{Chesi14,Miserev17}. Here, a $z\to -z$ asymmetry is induced by the vector potential which, for an in-plane magnetic field and using the Landau gauge, has the form ${\bf A}=(B_y z, -B_x z,0)$. A detailed analysis of these effective spin-orbit interactions, including estimates of various coupling coefficients and the leading corrections beyond the spherical approximation,  can be found in  \cite{winkler2003spin,Coish20,Culcer21} (and references therein). 

Regardless of their origin, all coupling coefficients in Eq.~(\ref{Hsoc}), and not only $\alpha_R$, are  strongly affected by the confinement potential $V(z)$. This becomes clear from their perturbative expressions, which depend on matrix elements and energy differences between 2D subband states. In general, the energy denominators of the perturbative expansion are on the order of the heavy-hole/light-hole splitting $\Delta_{\rm LH}$. Since $\Delta_{\rm LH}$ ranges from a few meV to several tens of meV, much less than the bulk band gap, spin-orbit interactions of holes are generally stronger than for $s$-type conduction electrons. Furthermore, it may be expected that a large value of $\Delta_{\rm LH}$ (due to, e.g., a thin well or large strain) will suppress spin-orbit interactions. This general tendency is true, but must be carefully weighted against the strength of the inversion-symmetry breaking mechanism. An interesting example is the non-monotonic dependence of $\alpha_R$ on the gate electric field, which was predicted theoretically \cite{Winkler00rashbaholes} and observed experimentally \cite{Winkler04}. While a moderate gate field enhances the spin-orbit coupling $\alpha_R$, due to the stronger asymmetry of $V(z)$ and larger heavy-hole to light-hole matrix elements, the limit of large fields is dominated by the growth of $\Delta_{\rm LH}$, which eventually leads to a decrease of $\alpha_R$. A different conclusion is found for the pseudospin-electric coupling contribution to $\beta_D$ which, despite being a Desselhaus term, is proportional to the applied gate field as well. Here, the growth of off-diagonal matrix elements was found to dominate over the energy denominators, leading to a monotonic dependence with gate field \cite{Coish20}.

\emph{Nanowires.} The case of one-dimensional systems is less straightforward to discuss since, assuming a confining potential $V(y,z)$ (with holes free to move along $x$), a factorization analogous to the 2D case does not happen at $k_x=0$. It is therefore necessary to consider in detail how all values of $J_z$ enter the low-energy eigenstates. Nevertheless, several general points discussed above are still valid. A finite spin-orbit coupling in the 1D subbands can be generated either by lack of bulk inversion symmetry or through an asymmetric external gate field, e.g., described by a term $eV(z) =e E_z z$ in Eq.~(\ref{Henv_funct}). In the important case of Si and Ge nanowires, the former type (Dresselhaus spin-orbit interaction) is absent in the bulk, and the main spin-orbit mechanism is expected to be of the Rashba type. However, in contrast to quantum wells, where the Rashba spin-orbit coupling is cubic in momentum [see Eq.~\eqref{Hsoc}], for nanowires it depends linearly on $k_x$.

The different (cubic or linear) dependence can be related to the exact decoupling between heavy-holes and light-holes at zero subband momentum, which only happens in 2D. For a quantum well, we see in Eq.~(\ref{HLut}) that the terms with a linear dependence on the in-plane compnents $k_i$  (where $i=x,y$) are proportional to $\{ J_{i},J_z\}$, thus vanish in the $J_z=\pm 3/2$ subspace (i.e., at first-order in perturbation theory). These terms can only contribute to the effective spin-orbit interaction at higher-order, introducing a non-linear dependence on $k_i$\footnote{For 2D holes, a Rashba interaction linear in momentum is  not forbidden by symmetry \cite{winkler2003spin}. However, it requires the contribution of distant conduction bands (see the coupling coefficient $r_{51}^{7h7h}$, given in Section 6.3.3 of Ref.~\cite{winkler2003spin}), thus is a small effect which goes beyond the Luttinger Hamiltonian.}. In 1D the same argument does not hold: The $V(y,z)$ confinement potential will induce a finite mixing of heavy- and light-holes in the lowest subband, even if $k_x=0$. The terms of the Luttinger Hamiltonian proportional to $\{J_x,J_j\}$ (where $j=y,z$) can generate a Rashba spin-orbit interaction linear in $k_x$. 

Another simple calculation which illustrates the appearance of a linear term in 1D considers a potential $V(y,z)=V_s(z)+V_w(y)$, where the vertical confinement along $z$ is much stronger than the confinement in the lateral direction, thus justifying the use of an effective 2D Hamiltonian. To first approximation, neglecting spin-orbit interactions and taking ${\bf B}=0$, the lowest 1D subbands are spin degenerate and have orbital wavefuction $\propto \psi_0(y) e^{ik_x x}$, where $\psi_0(y) $ is determined by $V_w(y)$ and by the in-plane effective mass. Using the Rashba spin-orbit coupling as described in Eq.~(\ref{Hsoc}), we obtain an effective 1D Hamiltonian by computing expectation values with respect to the wavefunction along $y$ (note that $\bra{\psi_0} p_y\ket{\psi_0} =\bra{\psi_0} p_y^3 \ket{\psi_0}=0$, since $\ket{\psi_0}$ is a bound state):
\begin{align}\label{Hsoc1D}
H^{\rm 1D}_{R}\simeq i\alpha_R  \bra{\psi_0} (p_+^3\sigma_- - p_-^3\sigma_+)\ket{\psi_0} \simeq -6 \alpha_R \bra{\psi_0} p_y^2\ket{\psi_0} p_x \sigma_y + \ldots,
\end{align}
where we only show the linear term, which dominates at small values of $k_x$. Including the cubic terms predicts a crossing point at larger $k_x$, which has experimental consequences in transport \cite{Chesi11}. The above simple argument shows explicitly how the lateral confinement along $y$ changes the dependence on momentum from cubic (in 2D) to linear (in 1D). Equation~(\ref{Hsoc1D}) also indicates that the spin-orbit splitting grows for a tighter confinement along $y$, due to the increase of $\bra{\psi_0} p_y^2\ket{\psi_0} $. For actual nanowires, where the confinement energies along $z$ and $y$ are roughly equivalent, this simple approach is not applicable. Explicit calculations based on realistic geometries were carried on in detail, finding a `direct Rashba' spin-orbit interaction linear in momentum which can be extremely large \cite{Kloeffel_PRB2011,KloeffelPRB2018}. For nanowires based on Ge and Si, the theoretical spin-orbit coupling energies can reach the meV scale. As we will discuss in Sec.~\ref{sec:nanowires}, results from experiments in Ge/Si core/shell and hut-wire nanowires corroborate these theoretical predictions of a large spin-orbit coupling. In the literature, a spin-orbit coupling linear in momentum, e.g., with a dependence $\alpha_{\rm so} p_x \sigma_y$ as in Eq.~(\ref{Hsoc1D}), is often charachterized through the spin-orbit length $\lambda_{\rm so}= \hbar/m^*\alpha_{\rm so}$, where $m^*$ is the effective mass. 

\subsection{Spin-orbit coupling and hole-spin qubits}\label{sec:SOC_qdots}

We now turn to quantum dots where, for a single hole confined in an external potential, the ground state is a degenerate Kramers doublet at ${\bf B}=0$.
These doublets, which are split in energy by an external magnetic field, form a natural basis for a qubit. The qubit states can be labeled using the pseudospin-1/2 states, $\ket{\Uparrow}$ and $\ket{\Downarrow}$.
It should be noted that, due to the presence of significant spin-orbit interactions, the states $\ket{\Uparrow}$ and $\ket{\Downarrow}$ cannot be decomposed into a product of spin and orbital degrees of freedom. This fact has important consequences on the properties of the effective $g$-tensor, electric manipulation of spin states, as well as spin relaxation and decoherence.

\emph{The $g$-factor of single holes} We start by considering one of the most basic properties relevant for spin qubits: the Zeeman splitting induced by an external magnetic field.
For holes, this splitting is found to depend on the direction of the magnetic field and on the details of the confinement potential.
In large lateral quantum dots formed starting from a 2DHG, where the extent of the wavefunction is much larger in the $xy$ plane than along the $z$ direction, the lowest-energy states can be well approximated by pure heavy holes.
Applying time inversion to the Zeeman term of Eq.~(\ref{Henv_funct}), and projecting it to the heavy-hole subspace gives:
\begin{equation}\label{Hzeeman}
H_Z \simeq -\left(3\kappa +\frac{27}{4}q \right)\mu_B B_z \sigma_z -\frac{3}{2} q \mu_B (B_x \sigma_x - B_y \sigma_y),
\end{equation}
where holes with $J_z=\pm 3/2$ are mapped to the $\sigma_z = \pm 1$ pseudospin states. The form of the transverse-coupling term ($\propto q$) reflects the fact that the zincblende and diamond lattices do not have 90-degree rotational symmetry (this term
changes sign under 90-degree rotations). Since $\kappa  \gg q$ (for example, $\kappa \simeq 1.2$ and $q \simeq 0.01$ in GaAs), the Zeeman splitting predicted by Eq.~(\ref{Hzeeman}) is strongly anisotropic. It is important to note, however, that Eq.~(\ref{Hzeeman}) neglects contributions from the admixture of light holes and heavy holes.
These contributions can be important and must therefore be included in general.

Interestingly, for the $z$-direction, a finite correction to $g_z \simeq  6\kappa$ survives even in the theoretical limit of an infinitely thin well ($L \rightarrow 0$), when the admixture of light-hole states becomes negligible \cite{Wimbauer94,Durnev13,Franceschi13}. The finite correction arises in second-order perturbation theory, due to a cancellation between the large gap ($\Delta_{\rm LH} \propto 1/L^{2}$) and the squared transition matrix elements between heavy and light holes (which formally diverge as $1/L$). As the gaps between subbands and all transition matrix elements depend on the detailed confinement potential $V(z)$, the value of $g_z$ can be modulated by a top gate and a large variation between $g_z \simeq 1$ and $2.5$ was reported for the SiGe self-assembled quantum dots of Ref.~\cite{Franceschi13}. The modulation of $g_z$ with gate voltage is a well-documented effect for spin qubits (e.g., in InGaAs \cite{Warburton15} and GaAs \cite{Sachrajda19} quantum dots), and we will further discuss it in Sec.~\ref{sec:experiments}. Similarly, the in-plane $g$-factors can be strongly affected by the confinement potential, as shown by the following simple example. We consider again a quasi-2D system and assume an elliptic confinement with principal axes along $x$ and $y$. Treating the $\gamma$ term of Eq.~(\ref{Hsoc}) as a perturbation, and taking the average over the orbital wavefunction, leads to:
\begin{equation}\label{delta_gx}
\delta H_Z \simeq 2\gamma (\langle p_x^2 \rangle - \langle p_y^2 \rangle )(B_x \sigma_x + B_y \sigma_y) .
\end{equation}
In elliptical dots, where $\langle p_x^2 \rangle \neq  \langle p_y^2 \rangle$, Eq.~\eqref{delta_gx} modifies the in-plane $g$-factors. We see that the change of $g_{x,y}$ can be controlled through the $z$ confinement, entering the $\gamma$ coupling, as well as the $xy$ potential, affecting $\langle p_{x,y}^2 \rangle \propto \omega_{x,y}$. Since Eq.~(\ref{Hzeeman}) gives small values of $g_{x,y}$, corrections like Eq.~(\ref{delta_gx}) can be particularly relevant. A good example is a recent measurement in a Ge double quantum dot \cite{Katsaros22}. While $3q \simeq 0.18$ in Ge, the measured in-plane $g$-factors were $\simeq 0.26$ and $-0.16$ for the left and right dot, respectively. A detailed theoretical analysis of the $g$-factor, starting from the Luttinger Hamiltonian [Eq.~\eqref{HLut}], finds a contribution $\propto \omega_x -\omega_y$, which is in agreement with Eq.~(\ref{delta_gx}). 
The interpretation given in Ref.~\cite{Katsaros22} is that, since the two quantum dots are elliptical and have major axes which are approximately perpendicular to each other, the corrections to the in-plane $g$-factors can be comparable or larger than $3q$.
Moreover, these corrections will have opposite sign for the two dots.
We note that while the effect caputred by Eq.~(\ref{delta_gx}) is only relevant for elliptical quantum dots, other spin-orbit couplings [beyond the $\gamma$ term of Eq.~\eqref{Hsoc}] can lead to a gate-dependent $g_x$ (as well as $g_z$) even for circular quantum dots \cite{Veldhorst22theory}. 

Similar features are found for the $g$-factor of nanowires. In Ge/Si core/shell nanowires, the $g$-factor can be strongly quenched along the direction of the nanowire \cite{Loss13,Zwanenburg16}. A recent detailed analysis of $g$-factors in a silicon nanowire device, together with the dependence on various gate potentials, can be found in Ref.~\cite{Franceschi22b}. A thorough characterization of the gate dependence and anisotropy is important to determine suitable sweet spots, at which the hole-spin qubit is less sensitive to charge noise \cite{Culcer21npj,Loss21b,Veldhorst22theory,Franceschi22b}.

\emph{Electric-dipole spin resonance.} One of the main advantages of hole-spin qubits is the potential for electrical manipulation which, as we will see in Sec.~\ref{sec:experiments}, has been demonstrated with impressive performance. Electric-dipole spin resonance (EDSR) relies on a purely electrical drive coupling the hole-spin states. We consider a uniform alternating electric field, ${\bf{E}}(t)= E(t)\hat{x}$ and the associated potential $eE(t)x$.
The basic mechanism for EDSR is a non-vanishing transition element $\bra{\Uparrow} eE(t)x \ket{\Downarrow} \neq 0 $, which is generally allowed when the Zeeman-split levels are not pure spin states. The combination of spin-orbit coupling and electric field behaves as a net magnetic field, and purely electrical spin rotations can be accomplished when the frequency of the electric field matches the Zeeman splitting between the qubit states.

More detailed considerations are necessary to establish the effectiveness of the electrical drive, as this depends on the geometry of the dot and the type of spin-orbit interaction. Some intuition may be gained from discussing spin-orbit coupling as a perturbation. In this language we see that, unlike electron-spin resonance (ESR, its purely magnetic counterpart), EDSR involves orbital excited states. This is because in a quantum dot the expectation value of the spin-orbit interaction within any one level vanishes, but it causes spin-flip transitions between levels. In other words, its matrix elements are purely off-diagonal.  It is simplest to assume a parabolic confinement potential for a circular quantum dot, with eigenstates given by the well-known Fock-Darwin states and where the qubit lies in the orbital ground state, labeled by the orbital quantum number $n=0$. The Dresselhaus terms of Eq.~(\ref{Hsoc})  connect the ground state $n = 0$ to the first orbital excited state $n = 1$, and the virtual transition is accompanied by a spin-flip. Since the electric potential $eE(t)x $ also connects the states with $n = 0$ and $n = 1$, but without an associated spin flip, virtual transitions from $n = 0$ to $n = 1$ and back appear in second order perturbation theory and couple opposite-spin states. For a circular dot, because of the presence of the electrical potential $eE(t)x$, the virtual transitions must involve the first orbital excited state, thus not every form of spin-orbit coupling leads to EDSR. Importantly, in Eq.~(\ref{Hsoc}) the Rashba term $\propto \alpha_R$ does \textit{not} yield EDSR for a circular dot, since it connects $n=0$ to $n=3$. Thus, it does not contribute to second-order perturbation theory in conjunction with $eE(t)x$. In the seminal paper that introduced EDSR for holes \cite{Loss07} it is the comparatively weaker interaction $\propto \beta_3$ that rotates the spin (linear-$k$ terms were not included in Ref.~\cite{Loss07}). If the linear Dresselhaus spin-orbit coupling dominates, the detailed analysis of EDSR becomes analogous to the case for conduction electrons \cite{Loss06}. 

As the $\alpha_R$ term turns out to be ineffective for circular dots, an interesting question concerns the mechanism of EDSR for materials where Dresselhaus terms are absent in the bulk, such as Si and Ge. In this case, a sizable $\beta_D$ interaction may be induced through interface inversion asymmetry \cite{Durnev14} or through a finite dipolar coupling \cite{Coish20}. Even in the absence of Dresselhaus interactions, the Rashba spin-orbit coupling leads to EDSR if the 2D dot deviates from circular symmetry (and, by extension, in nanowires). Furthermore, the leading cubic-symmetry correction to the Rashba interaction, obtained by going beyond the spherical approximation of $H_L$, has the same formal dependence of the term $\propto \beta_3$ and has been shown to allow for efficient EDSR spin manipulation in Ge quantum dots  \cite{Culcer21}. 

\emph{Decoherence.} In the earliest days of quantum-dot spin qubits the idea was prevalent that quantum-dot spins ought to exhibit strong coherence properties because the diagonal matrix elements of the spin-orbit interaction vanish (leading to a vanishing pure-dephasing rate \cite{Loss04,bulaev2005spin}). Over time it has been recognized that off-diagonal  matrix elements of the spin-orbit coupling have a very strong effect on spin coherence times. Moreover, in contrast to EDSR where only some spin-orbit couplings have a non-vanishing contribution, all spin-orbit couplings can contribute to decoherence. In current hole-spin devices, spin-orbit coupling is usually regarded as the main decoherence engine since, as explained in detail in the next section, the $p$-character of their orbital wave functions ensures the contact hyperfine interaction vanishes, leaving only the weaker orbital and dipolar hyperfine couplings.

It is generally believed that relaxation is associated primarily with phonons, while the main dephasing mechanisms are phonons and background charge fluctuations. The electrical potentials describing phonons and charge noise connect the ground state to all higher orbital excited states.
Consequently, phonons and charge noise combined with spin-orbit coupling affect the qubit subspace (in second order perturbation theory). The spin-flip matrix elements induced by phonons are analogous to those enabling EDSR, but become now responsible for qubit relaxation. For holes, the relaxation time $T_1$ due to the phonon mechanism has been shown to behave as $T_1 \propto 1/B^5$ for cubic Dresselhaus, and $T_1 \propto 1/B^9$ for Rashba dominated spin-orbit interaction \cite{bulaev2005spin}. Recent measurements of $T_1$ in a GaAs quantum dot find a dependence close to $\propto 1/B^5$, favoring the first scenario \cite{Sachrajda19relax}. Note, however, that linear spin-orbit coupling gives a $\propto 1/B^5$ dependence  as well \cite{Loss04}.
Thus, the data might be compatible with a dominant $\beta_D$ term \cite{Coish20}, see Eq.~(\ref{Hsoc}).  For quantum dots based on Si and Ge, theoretical studies of the spin lifetime are also available \cite{Loss13,Culcer21npj}.

As discussed above in some detail, the $g$-factor depends on the electrostatic environment of the hole spin, and this dependence provides a natural decoherence mechanism.
Like gate potentials, electric fluctuations induced by phonons and charge noise can modify the $g$-factor and lead to dephasing.
For this reason, hole-spin dephasing is often related to the spectral properties of classical charge noise (see for example Refs.~\cite{Huthmacher18,Veldhorst21,Franceschi22b}). We also note that the EDSR strength is linear in the spin-orbit coupling, while the relaxation and dephasing rates are quadratic. Consequently, the existence of a favorable range of spin-orbit coupling strengths, where fast EDSR is possible with relatively slow relaxation and dephasing rates, is generally expected. 

\subsection{Hyperfine interaction}\label{sec:hyperfine}

Understanding the hyperfine coupling of electron and hole spins to an uncontrolled environment of nuclear spins is essential for a broad class of qubits based on quantum dots \cite{Schliemann03,Coish09,Chekhovich13a,Warburton13review}. This is especially true for \emph{electron} spins in quantum dots based on III-V semiconductors since every naturally occurring isotope of every group III and group V element carries a finite nuclear spin \cite{Schliemann03}, and the conduction band of III-V materials is predominantly $s$-like, resulting in a strong contact interaction. The significant dephasing induced by nuclear spins in early III-V spin qubits was a main motivation to develop alternative platforms, and hole spins in III-V or group IV quantum dots show great promise in this respect. Since the contact hyperfine coupling vanishes for the predominantly $p$-like states of the valence band, the influence of nuclear spins is expected to be generally weaker for holes \cite{Loss08decoherence, Chekhovich13b, Vidal2016, Warburton16}, and the anisotropy of the hole-spin hyperfine interaction admits motional averaging that is not possible for an isotropic contact interaction \cite{Loss08decoherence, Wang12, Chesi14hf, Wang15}.

A direct approach to suppress decoherence from nuclear spins is through isotopic purification, which is possible for devices based on Si and Ge. However, even a trace amount (800 p.p.m.) of $^{29}$Si in some isotopically enriched $^{28}$Si devices has been shown to significantly influence electron-spin dynamics \cite{Zhao19, Hensen20}. As we will discuss, the hyperfine interaction in the conduction band of Si has a strength comparable to values predicted for the valence band of Si and Ge (see Table~\ref{tab:hfparamsVB}), suggesting that hyperfine interaction can affect the hole-spin dynamics in many practically relevant scenarios. As effective strategies to cope with charge noise are being developed, the role of hyperfine interaction will presumably grow in importance for Si and Ge hole-spin devices. In fact, coherence times compatible with hyperfine interaction timescales were already reported in Si FinFET devices~\cite{Camenzind22}. Establishing a solid understanding of the hyperfine coupling in each material, and for each type of device, could also allow for alternative strategies, where the hyperfine coupling is exploited as a resource (e.g., by storing quantum information in the nuclear degrees of freedom \cite{Taylor03,Hensen20}), rather than suppressed as an unwanted source of decoherence. 

\emph{Microscopic theory.} For an electron moving in the electric and magnetic fields generated by a collection of spinful nuclei (spins $\left\{\mathbf{I}_l\right\}$ at sites $\left\{\mathbf{r}_l\right\}$), a non-relativistic expansion of the Dirac equation leads to a sum of three terms that couple the electron orbital ($\mathbf{r}, \mathbf{p}$) and spin ($\mathbf{S}$) degrees of freedom to  the nuclear spin ${\bf I}_l$: 
\begin{equation}
\mathcal{H}_\mathrm{hf}=\sum_l\left(\mathcal{H}^l_\mathrm{c}+\mathcal{H}^l_\mathrm{dip}+\mathcal{H}^l_\mathrm{orb}\right).
\end{equation}
The first contribution is the Fermi contact interaction, which determines the hyperfine interaction in the conduction band. The explicit expression is (with $\hbar=1$):
\begin{equation}
\mathcal{H}_\mathrm{c}^l=\frac{\mu_0}{4\pi}\frac{8\pi}{3}|\gamma_S|\gamma_l\delta_\mathrm{T}(\mathbf{r}-\mathbf{r}_l)\mathbf{S}\cdot\mathbf{I}_l,   
\end{equation} 
where $\gamma_S=-2\mu_\mathrm{B}$ is the electron gyromagnetic ratio, $\gamma_l$ is the nuclear gyromagnetic ratio, and 
$ \delta_\mathrm{T}(\mathbf{r})=\frac{1}{4\pi r^2}\frac{d f_\mathrm{T}(r)}{dr}$ is a localized function. This function is written in terms of $f_\mathrm{T}(r) = r/(r+r_\mathrm{T}/2)$, where $r_\mathrm{T}=Z\alpha^2 a_B$ is the Thomson radius for a nucleus of core charge $Z$, with $\alpha\simeq 1/137$ being the fine structure constant and $a_B$ the Bohr radius. Since the Fermi contact interaction vanishes in the valence band, the hyperfine interaction of hole spins is entirely due to the dipolar and orbital couplings. The corresponding expressions read:
\begin{equation}
\mathcal{H}_\mathrm{dip}^l=\frac{\mu_0}{4\pi}|\gamma_S|\gamma_l\mathbf{S}\cdot\overleftrightarrow{D}(\mathbf{r}-\mathbf{r}_l)\cdot\mathbf{I}_l,\label{eq:Hdip}
\end{equation}
where the dipolar tensor has elements $D_{\alpha\beta}({\bf r})=\frac{3r_\alpha r_\beta-r^2\delta_{\alpha\beta}}{r^5}f_T(r) $, and:
\begin{equation}
\mathcal{H}_\mathrm{orb}^l=\frac{\mu_0}{4\pi}|\gamma_S|\gamma_l\frac{1}{|\mathbf{r}-\mathbf{r}_l|^3}f_\mathrm{T}(|\mathbf{r}-\mathbf{r}_l|)\mathbf{L}_l\cdot\mathbf{I}_l,\label{eq:Horb}
\end{equation}
where $\mathbf{L}_l=\left(\mathbf{r}-\mathbf{r}_l\right)\times \mathbf{p}$ gives the electron orbital angular momentum about site $l$.  Note that in the extreme non-relativistic limit, $r_\mathrm{T}\to 0$, $f_\mathrm{T}(r)\to 1 $ and $\delta_\mathrm{T}(\mathbf{r})\to\delta(\mathbf{r})$ reduces to a 3D Dirac delta function.  Provided the nonrelativistic Schr\"odinger equation is used consistently to solve for the electronic states, this limit can still lead to accurate results for the hyperfine parameters of small-$Z$ nuclei (with the small parameter $Z\alpha\ll 1$).  However, the results become progressively quantitatively less accurate for large $Z$ \cite{Pyykko73}.  Although it may be appropriate to use solutions to the non-relativistic Schr\"odinger equation to derive hyperfine couplings from the extreme non-relativistic form of the hyperfine interactions ($r_\mathrm{T}\to 0$), it is absolutely necessary to consistently include $r_\mathrm{T}\ne 0$ if relativistic solutions for the electronic states are found, since these relativistic solutions may vary rapidly on the scale $r\sim r_\mathrm{T}$ \cite{Philippopoulos20}.

To make use of the microscopic Hamiltonian $\mathcal{H}_\mathrm{hf}$ for a qubit described by two energetically well-isolated states $\ket{\Uparrow},\ket{\Downarrow}$, we derive an effective Hamiltonian via the projector $P=\sum_{s=\Uparrow,\Downarrow}\ketbra{s}{s}$:
\begin{equation}
    H_\mathrm{hf} = P\mathcal{H}_\mathrm{hf}P = \sum_l\left( \mathbf{s}\cdot\overleftrightarrow{A}_l\cdot\mathbf{I}_l+\mathbf{B}_l\cdot\mathbf{I}_l\right).
\end{equation}
Here, we have introduced a pseudospin $\mathbf{s}$ that acts in the space of qubit states [e.g., $s_z=(\ket{\Uparrow}\bra{\Uparrow}-\ket{\Downarrow}\bra{\Downarrow})/2$].  For holes, the hyperfine tensor elements $A_l^{\alpha\beta}$ and the field components $B_l^\alpha$ are found from matrix elements of Eqs.~(\ref{eq:Hdip}) and (\ref{eq:Horb}) with respect to the qubit states.  When $\ket{\Uparrow},\ket{\Downarrow} $ describes a Kramers doublet (two states related by time-reversal), $\mathbf{B}_l=0$ identically, which follows from the fact that all terms in $\mathcal{H}_\mathrm{hf}$ are odd under time reversal \cite{Philippopoulos20}. Furthermore, as discussed, the $\ket{\Uparrow},\ket{\Downarrow}$ states of a hole qubit established in a quantum dot  are commonly described through the envelope function approximation. Explicitly, we have ($s=\Uparrow,\Downarrow$):
\begin{equation}\label{eq:QubitEFA}
    \braket{\sigma \mathbf{r}}{s} \simeq \sqrt{v_0} \sum_{j_z=-3/2}^{3/2}F^{(s)}_{j_z}(\mathbf{r})u_{j_z} (\mathbf{r},\sigma),
\end{equation}
where $u_{j_z} (\mathbf{r},\sigma)$ are the lattice-periodic Bloch amplitudes at crystal momentum ${\bf k}=0$.  The Bloch amplitudes are normalized to $\sum_{\sigma}\int_\Omega d^3 r |u_{j_z}(\mathbf{r},\sigma)|^2=n_a$ over the unit cell volume $\Omega$, where $n_a$ is the number of atoms in the unit cell and $v_0=\Omega/n_a$ is the volume per atom. The envelope functions satisfy the normalization condition $\sum_{j_z}\int d^3 r|F^{(s)}_{j_z}({\bf r})|^2=1$. When the qubit states can be expressed as in Eq.~(\ref{eq:QubitEFA}), the hyperfine tensor elements can be calculated in a straightforward two-step process.  First, a microscopic description of the electronic structure for the translationally invariant crystal gives an estimate for the Bloch amplitudes, $u_{j_z} (\mathbf{r},\sigma)$, allowing for a calculation of the material-dependent (but device-independent) part of the hyperfine tensor.  Second, modeling of the device or nanostructure may give a reasonable description of the envelope functions, allowing for the final device-dependent hyperfine tensor to be found in terms of the device-independent material parameters. 

\emph{Hyperfine parameters of holes.} Carrying out the above procedure, the device-independent hyperfine interaction with nucleus of isotope $i_l$ at site $l$ leads to the following contribution that can be added to the effective Hamiltonian [Eq.~(\ref{Henv_funct})] for the envelope functions  \cite{Chekhovich13b,Philippopoulos20}:
\begin{equation}\label{eq:Hl}
\mathrm{H}^l  =  \left[\left(\frac{1}{3}A_\parallel^{i_l}-\frac{3}{2}A_\perp^{i_l}\right)\mathbf{J}\cdot\mathbf{I}_l+\frac{2}{3}A_\perp^{i_l}\boldsymbol{\mathcal{J}}\cdot\mathbf{I}_l\right] v_0 \delta({\bf r}-{\bf r}_l).
\end{equation}
Here, the Dirac delta function bears no relation to the Fermi contact interaction, since Eq.~(\ref{eq:Hl}) is only determined by the dipolar and orbital terms. The presence of the delta function reflects the fact that the envelope functions are approximately constant over the scale of a unit cell, thus the hyperfine interaction acts on the 4-component spinor $\left(F^{(s)}_{3/2}(\mathbf{r}_l),F^{(s)}_{1/2}(\mathbf{r}_l),F^{(s)}_{-1/2}(\mathbf{r}_l),F^{(s)}_{-3/2}(\mathbf{r}_l)\right)^T$ at the position of the nuclear spin.  Notably, Eq.~\eqref{eq:Hl} shows that the hyperfine tensor for a general hole-spin qubit (any linear combination of heavy holes and light holes) can be expressed in terms of just two material parameters per isotope: $A_\parallel^i,\,A_\perp^i$. The dependence on the angular momentum matrices $J_i$ takes the same form as in the Zeeman term, see  Eq.~(\ref{Henv_funct}), as it is determined by the tetrahedral point-group symmetry of the zincblende lattice. In particular, a nonzero value of $A_\perp^i$ is due to the hybridization of $p$ and $d$ orbitals induced by the crystal field \cite{Chekhovich13b,Philippopoulos20}. Specializing to an ideal heavy-hole qubit and, assuming  for simplicity a spin-independent envelope function $F(\mathbf{r})$, one obtains \cite{Chekhovich13b,Philippopoulos20}:
\begin{equation}\label{eq:H_HeavyHole}
    H_\mathrm{hf}^\mathrm{HH} =\sum_lv_0\left|F(\mathbf{r}_l)\right|^2\left[A^{i_l}_{\parallel}s_zI^{z}_l+A^{i_l}_{\perp}\left(s_xI^{x}_l-s_yI^{y}_l\right)\right].
\end{equation}
Similarly, projecting ${\rm H}^l$ to the light-hole subspace gives \cite{Chekhovich13b,Philippopoulos20}:
\begin{equation}\label{eq:HLightHole}
    H^\mathrm{LH}_{\mathrm{hf}} =\frac13 \sum_lv_0\left|F(\mathbf{r}_l)\right|^2\left[\left(A^{i_l}_{\parallel} - 4 A^{i_l}_{\perp}\right) s_zI^{z}_l+ \left(2A^{i_l}_{\parallel} + A_{\perp}^{i_l}\right)\left(s_xI^{x}_l+s_yI^{y}_l\right)\right], 
\end{equation}
which corrects the expression given in Appendix D of Ref.~\cite{Philippopoulos20}. Since typically $A_\perp \ll A_\parallel$, the above limiting cases lead to hyperfine interactions of very different character. For heavy holes, the Ising term ($\propto s_zI^z_l$) dominates for all nuclear species listed in Table \ref{tab:hfparamsVB}. For light holes instead, the in-plane coupling ($\propto s_xI^x_l, s_yI^y_l$) is always larger. Interestingly, for the Ga site of GaAs an approximate cancellation of the Ising term occurs in Eq.~(\ref{eq:HLightHole}), based on the density-functional theory results of ~Ref.~\cite{Philippopoulos20} (collected in Table \ref{tab:hfparamsVB}).  Due to these different behaviors, a significant heavy-hole/light-hole mixing will result in a nontrivial form of the hyperfine interaction. For spin qubits based on Si and Ge (where the effect of $d$-orbital hybridization is small, see Table~\ref{tab:hfparamsVB}), a carefully chosen confinement geometry and magnetic field setting can lead to a qubit that is comparatively insensitive to nuclear-field fluctuations \cite{Loss21a}. On the other hand, for self-assembled quantum dots the effect of heavy-hole/light-hole mixing was found to be smaller than $d$-orbital hybridization \cite{Chekhovich13b,Cywinski19}.            

\begin{table}
\centering
\begin{tabular}{l|c|cc}
\hline\hline
				& (electrons) 					&\multicolumn{2}{c}{(holes)}\\
  isotope ($i$) & $A^i$ ($\mu e\mathrm{V}$) & $A^i_{\parallel}$ ($\mu e\mathrm{V}$) & $A^i_{\perp}$ ($\mu e\mathrm{V}$) \\
   \hline
   $^{69}\mathrm{Ga}$ in GaAs & $74$ &  $1.4$ & $0.35$\\   
   $^{71}\mathrm{Ga}$ in GaAs & $94$ & $1.7$ & $0.45$ \\
   $^{75}\mathrm{As}$ in GaAs & $78$ & $11$ & $0.02$ \\ 
   $^{29}\mathrm{Si}$ in silicon & $-2.4$ &   $-2.5$ & $-0.01$  \\  
	 $^{73}\mathrm{Ge}$ in germanium & - &   $-1.1$ & $-0.02$  \\
\hline\hline
\end{tabular}
\caption{Theoretical hyperfine parameters calculated in Ref.~\cite{Philippopoulos20} and  (in the case of $^{73}\mathrm{Ge}$) Ref.~\cite{PerryPhD}, using first-principles density functional theory.  }\label{tab:hfparamsVB} 
 \end{table}

Table \ref{tab:hfparamsVB} shows a summary of all hyperfine parameters found within density functional theory in Ref.~\cite{Philippopoulos20}, as well as the valence-band hyperfine parameters obtained in Ref.~\cite{PerryPhD} for germanium. For reference, we also list the hyperfine parameters obtained for the conduction band, which compare well with early estimates in GaAs \cite{Paget77} and Knight shift measurements (see Ref.~\cite{Philippopoulos20} for details).  Remarkably, the strength of the hyperfine coupling for holes in silicon is found to be comparable to the strength of the hyperfine coupling for electrons in a silicon conduction band.  This is a consequence of $s$-$p$ hybridization in the conduction band of silicon, resulting in an overall reduction in the contact interaction, rather than any enhancement of the hole hyperfine coupling. The hyperfine parameters in the valence band of Ge are comparable in magnitude to those for Si, and show a similar high degree of anisotropy ($A^i_{\parallel}\gg A^i_{\perp}$). An important next step will be to experimentally measure the hole hyperfine parameters listed in Table \ref{tab:hfparamsVB}.  Until now, the transverse ``flip-flop'' components of the hyperfine tensor have been largely inaccessible, with most measurements focusing on the longitudinal Overhauser field \cite{Fallahi10,Chekhovich11, Vidal2016, Warburton16}.  Hole spin echo envelope modulations (HSEEM) \cite{Philippopoulos19} should be highly sensitive to these transverse terms and may provide a path to extracting these parameters in the near future.

\section{Experimental platforms}\label{sec:experiments}

Hole-spin qubits form a varied landscape, with a number of experimental platforms being actively developed. In the past few years planar Ge quantum dots have advanced rapidly and currently excel in several critical aspects, such as fidelity of single-spin manipulation and the demonstration of universal logic in larger ($2\times 2$) qubit arrays. However, as we will emphasize, each implementation brings along distinct advantages and interesting prospects of future progress.

\subsection{Self-assembled quantum dots (III-V)} 
\begin{figure}
\centering
\includegraphics[width=0.9\textwidth]{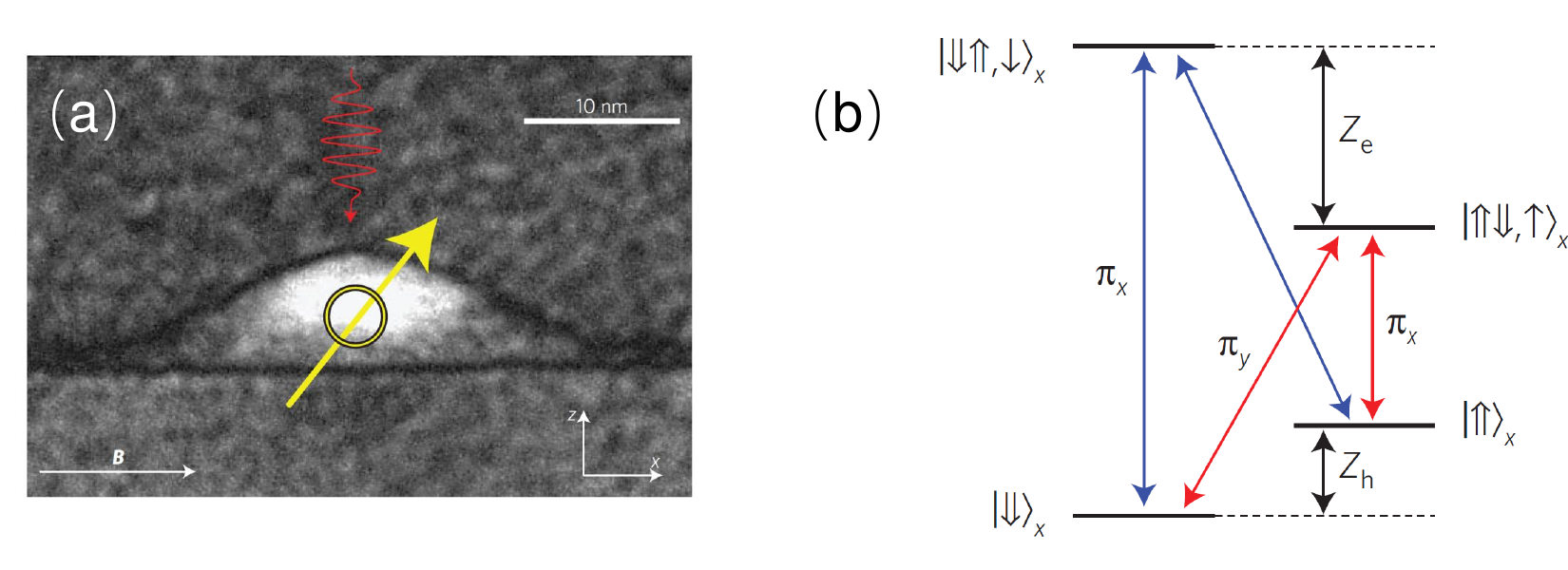}
\caption{(a) A typical self-assembled quantum dot in the Voigt configuration. (b) Level structure, including single hole states and positively charged trions. $Z_h$ ($Z_e$) is the hole (electron) Zeeman splitting and allowed optical transition, with linear polarization along $x$ ($\pi_x$) or $y$ ($\pi_y$), are also indicated. Both panels are adapted from Ref.~\cite{Warburton16}.}
\label{self-assembl_QD}
\end{figure}

Coherent spin manipulation of single holes was first demonstrated in 2011, based on self-assembled InGaAs/GaAs quantum dots \cite{Yamamoto11,Greilich11}. Figure~\ref{self-assembl_QD}(a) shows a typical self-assembled quantum dot, which is obtained upon epitaxial deposition of a thin `wetting layer' of InAs on a GaAs substrate~\cite{Warburton13review}. Due to the lattice mismatch, small InAs islands form spontaneously. The quantum dots are subsequently capped by GaAs which, thanks to the smaller bandgap of InAs, provides three-dimensional confinement for both electrons and holes. The small size of the dot results in large orbital energies $\sim 10$~meV, and allows for operation at liquid-helium temperatures instead of the mK regime. 
 
\emph{Single-spin manipulation.} A central feature of these spin qubits is that they are optically addressable~\cite{Warburton13review}. Single-spin manipulation can be realized with optical techniques on an ultrafast (ps) timescale and, to achieve this, the system is usually operated in the Voigt geometry, i.e., an in-plane magnetic field is applied perpendicular to the optical axis, see Fig.~\ref{self-assembl_QD}(a). As explained in Sec.~\ref{sec:theory}, see Eq.~(\ref{eq:H_HeavyHole}) and Table~\ref{tab:hfparamsVB}, the Overhauser field acting on heavy-holes is predominantly along $z$. As a consequence, the choice of a Voigt geometry also greatly improves coherence, by suppressing the effect of hyperfine interaction \cite{Loss08decoherence, Warburton16}. The relevant states and optical transitions are illustrated in Fig.~\ref{self-assembl_QD}(b) and involve single-hole spin states $\ket{\Uparrow},\ket{\Downarrow}$ and trion states $\ket{\Uparrow\Downarrow\uparrow},\ket{\Uparrow\Downarrow\downarrow}$, where an additional electron-hole pair is present in the quantum dot. By driving resonantly the $\ket{\Downarrow}\leftrightarrow\ket{\Uparrow\Downarrow\downarrow}$ transition, the state $\ket{\Uparrow}$ can be initialized with high fidelity through optical pumping \cite{Warburton08}. Readout can be performed in a similar manner, by monitoring the population of $\ket{\Downarrow}$ through the fluorescent light emitted by $\ket{\Uparrow\Downarrow\downarrow}$. As the transition $\ket{\Uparrow}\leftrightarrow\ket{\Uparrow\Downarrow\uparrow}$ remains undriven, $\ket{\Uparrow}$ does not give a readout signal. Finally, spin manipulation can be achieved through short (a few ps) off-resonant optical pulses \cite{Yamamoto11,Greilich11,Godden12}. As specified in Fig.~\ref{self-assembl_QD}(b),  in the Voigt geometry all four transitions between $\ket{\Uparrow},\ket{\Downarrow}$ and $\ket{\Uparrow\Downarrow\uparrow},\ket{\Uparrow\Downarrow\downarrow}$ are dipole-allowed for light with linearly polarization along $x$ or $y$. The selection rules are such that a circularly polarized broadband pulse will drive the stimulated Raman transition $\ket{\Uparrow} \leftrightarrow \ket{\Downarrow}$, thanks to the two available $\Lambda$-systems (for example, $\ket{\Uparrow}\leftrightarrow\ket{\Uparrow\Downarrow\downarrow}  \leftrightarrow \ket{\Downarrow}$). Such pulses implement effective Rabi oscillations between $\ket{\Uparrow} $ and $\ket{\Downarrow}$, with a rotation angle which depends on the applied power.
Combining the optically induced rotations (along $z$) with the free precession induced by the external magnetic field (along $x$), it is possible to achieve full control on the Bloch sphere. Recently, as an alternative to broadband pulses, a Raman laser technique was introduced for electron spins in InGaAs/GaAs quantum dots \cite{Atature19} and successfully applied to single holes \cite{Appel21,Appel22}.

\emph{Coherence.} These hole spin qubits display good coherence properties compared to their electron spin counterparts, where strong hyperfine interaction leads to dephasing times $T_2^*$ of only a few ns. Spectroscopic studies of single holes, based on coherent population trapping, are consistent with dephasing times of at least $100-500~$ns \cite{Warburton09,Warburton14,Warburton16}. However, shorter values are observed in pulsed experiment. For example, a typical $T^*_2 \sim 20$~ns was extracted from the decay of Ramsey fringes \cite{Greilich11}, and similar values were obtained from spin-flip Raman scattering \cite{Imamoglu16}. A non-monotonic dependence on the applied magnetic field was explicitly demonstrated in \cite{Huthmacher18}, with the longest coherence time $T_2^* \simeq 70$~ns occurring at $B_x =4$~T \cite{Huthmacher18}. Consistently with early reports \cite{Yamamoto11,Greilich11}, the high-field dephasing is attributed to to voltage fluctuations affecting the $g$-factor, while nuclear spins likely dominate noise at low fields \cite{Huthmacher18}. Coherence can be effectively prolonged through dynamical decoupling, giving $T^{\rm HE}_2 \sim 1-2~\mu$s  for the Hahn-echo decay time \cite{Yamamoto11,Huthmacher18}. A coherence time of $T_2^{\rm CP} \simeq 4~\mu$s was achieved through a Carr and Purcell (CP) decoupling sequence of $N_\pi = 8$ pulses \cite{Huthmacher18}. Further increasing $N_\pi$ is not beneficial, due to limitations in the $\pi$-pulse fidelity. Several recent experiments have reported a $\pi$-pulse fidelity around $F_\pi\simeq 90\%$ \cite{Huthmacher18,Appel21,Appel22}.

\emph{Entangling operations.} For self-assembled quantum dots, scaling-up of single qubits through fabrication is challenging. While quantum dot molecules of holes (formed in two vertically stacked dots) are actively studied \cite{Gammon12,Gammon19,Finley22} and, in fact, the coherent exchange interaction of two holes was demonstrated early on \cite{Greilich11}, an attractive alternative is to establish entanglement through the emitted photons. In 2016, an entangled state of two distant hole spins (5 m apart) was generated through an heralded protocol, based on the emission of Raman photons \cite{Delteil_2016}. Applying weak pulses resonant with the $\ket{\Downarrow}\leftrightarrow\ket{\Uparrow\Downarrow\downarrow} $ transition results in a small probability of spin flip, $\ket{\Downarrow}\rightarrow\ket{\Uparrow}$, accompanied by the emission of a single Raman photon. If two distant quantum dots initialized in the $\ket{\Downarrow}$ state are simultaneously driven in this manner, a Raman photon can be emitted by either of the dots (emission of two photons is highly unlikely). The experimental apparatus is carefully designed to erase the `which-path' information, such that detection of a single Raman photon heralds the desired entangled state, ideally of the form $(\ket{\Uparrow}\ket{\Downarrow}+ e^{-i\theta}\ket{\Downarrow} \ket{\Uparrow})/\sqrt{2}$ \cite{Delteil_2016}. In the actual experiment a fidelity $\sim 55~\%$ could be achieved \cite{Delteil_2016}.    

\emph{Hole spins as quantum emitters.} Self-assembled quantum dots are excellent quantum emitters of single photons \cite{Yamamoto02,Shields16,Senellart17}. A number of interesting protocols is based on this property, allowing to generate spin-photon entanglement as well as various classes of entangled photon states. A basic process is the same type of Raman scattering useful for remote entanglement. Upon application of a driving pulse on the `diagonal' transition $\ket{\Downarrow}\leftrightarrow\ket{\Uparrow\Downarrow\uparrow} $, see Fig.~\ref{self-assembl_QD}(b), Raman scattering with probability $p_1$ leads to a final state of the form $\sqrt{1-p_1}\ket{\Downarrow} \ket{0} + \sqrt{p_1}\ket{\Uparrow} \ket{1}$, where the states $\ket{0},\ket{1}$ refer to the emitted photon~\cite{Shields18PRX}. Crucially, the process where a photon is emitted but the hole spin returns to the initial state $\ket{\Downarrow}$ should be suppressed, which is achieved by making the ratio $C=\gamma_v/\gamma_d$ between the vertical/diagonal decay rates of the $\ket{\Uparrow\Downarrow\uparrow} $ trion as large as possible. While in free space  $C=1$, a Purcell enhancement of $\gamma_v$ by a factor of about 4 was achieved by embedding the quantum dot in a micropillar cavity \cite{Shields18QST,Shields18PRX}. Furthermore, by coupling the quantum dot to a photonic crystal waveguide, a large cyclicity $C=14.7$ could be demonstrated \cite{Appel21}. 

The above scheme can be extended to multiple pulses. A second pulse can further deplete $\ket{\Downarrow} \ket{0}$, leading to a state of the form $\sqrt{1-p_1-p_2}\ket{\Downarrow} \ket{00} + \ket{\Uparrow}(\sqrt{p_2}e^{i\phi} \ket{10} + \sqrt{p_1}\ket{01} )$. Here we indicate with $\ket{n_2n_1}$ the photon occupations in the temporal modes after the first ($n_1$) and second ($n_2$) pulse. The states $\ket{01}\equiv\ket{e}$ and $\ket{10}\equiv\ket{l}$ (`early' and `late' phonon, respectively) serve as a basis for a photonic time-bin qubit. An arbitrary time-bin state can be prepared by a suitable choice of $p_{1,2}$ and $\phi$, controlled by the intensities and relative phase of the two pulses \cite{Shields18PRX}. Continuing the process allows to generate time-bin encoded W-states, and sequences up to 4 pulses were implemented \cite{Shields18QST}. 

Such protocols can also take advantage of active spin-manipulation. A recent demonstration of spin-photon entanglement \cite{Appel22} starts from $(\ket{\Downarrow} + \ket{\Uparrow} )/\sqrt{2} $, obtained after applying a $\pi/2$ spin rotation to $\ket{\Downarrow}$. By driving the `vertical' transition  $\ket{\Uparrow}\leftrightarrow\ket{\Uparrow\Downarrow\uparrow} $, a state of the form $(\ket{\Downarrow} \ket{0} + \ket{\Uparrow} \ket{1})/\sqrt{2} $ is obtained. Afterwards, a $\pi$ rotation of the hole spin swaps $\ket{\Uparrow},\ket{\Downarrow} $ and a second excitation pulse is applied, with a controlled phase difference $\phi$. Ideally, the final result is a maximally entangled state $( e^{i \phi}\ket{\Uparrow} \ket{l}-\ket{\Downarrow} \ket{e})/\sqrt{2} $ between the hole-spin and the time-bin qubit. In practice, the state was obtained with a $F_{\rm Bell} \simeq 68\%$ fidelity, mainly limited by incoherent spin flips degrading the gate fidelity of the hole spin \cite{Appel22}. In principle, with additional rotations and excitation pulses, the protocol can be extended to generate GHZ and linear cluster states of multiple time-bin qubits~\cite{Appel22,Shields19}.

Photonic cluster states can also be generated following an early `photonic machine gun' proposal \cite{Lindner09}, which was recently implemented using a hole-spin \cite{Gershoni21}. At variance with previous discussions, this protocol is based on the optical selection rules of spin states $\ket{\Uparrow}_z,\ket{\Downarrow}_z$, i.e., with quantization axis along $z$. Furthermore, the in-plane magnetic field should be sufficiently small, allowing to neglect its effect during a fast process of optical excitation to the trion and subsequent decay. Under such conditions, linearly polarized resonant pulses induce the entanglement process $\ket{\Uparrow}=(\ket{\Uparrow}_z+\ket{\Downarrow}_z)/\sqrt{2} \to (\ket{\Uparrow}_z \ket{\sigma^-}+\ket{\Downarrow}_z \ket{\sigma^+})/\sqrt{2}$, where the photonic qubit is defined by the circular polarization ($\sigma_\pm$) of emitted photons. Each pulse adds a photon to the output state and synchronizing the pulses with $\pi/2$ spin rotations along $x$ (induced by the weak magnetic field) generates a one-dimensional photonic cluster state \cite{Lindner09}. The experimental protocol applies three excitation pulses and demonstrates significant improvements from a previous realization based on a dark exciton \cite{Gershoni16}. An extrapolated entanglement length of $\xi_{\rm LE} \simeq 10$ photons is obtained at $B=0.1$~T, substantially extending the previous $\xi_{\rm LE} \simeq 3.2$ \cite{Gershoni16}. Furthermore, relying on the hole spin allows to reach GHz photon generation rate with a much higher photon indistinguishability \cite{Gershoni21}.

\subsection{Planar quantum dots in GaAs}\label{sec:planar_GaAs}

Hole-spin qubits in III-V semiconductors can also be formed in gated structures, which offer better prospects for scalability through nanofabrication. Despite the early success of GaAs quantum dots hosting electron-spin qubits~\cite{Hanson07}, the corresponding hole-spin devices only witnessed significant progress in recent years (see \cite{Studenikin21review} for a detailed review). After developing more stable devices based on undoped heterostrucures, where the 2DHG is formed at a GaAs/AlGaAs interface using a top accumulation gate \cite{Hamilton10,Hamilton12}, the charge stability diagram of a GaAs double quantum dot could be mapped in 2014 down to the single-hole regime \cite{Tracy14}. 

At the moment of writing, hole-based double quantum dots in GaAs have been well characterized but spin manipulation is only at a preliminary stage, especially when compared to planar quantum dots in strained Ge quantum wells (see the next section). Besides fabrication issues, a disadvantage of GaAs is that every isotope of Ga and As has non-vanishing spin.
Therefore, in contrast to silicon and germanium nanostructures, the nuclear-spin environment cannot be eliminated via isotopic purification. Evidence of dynamic nuclear polarization was reported in Ref.~\cite{Studenikin21review} but, as the strength of the hyperfine interaction is much weaker than for electron spins, nuclear spins should not prevent substantial progress in the near future.   In this section we will also introduce several basic concepts relevant for all gated quantum dots.

\begin{figure}
\centering
\includegraphics[width=0.8\textwidth]{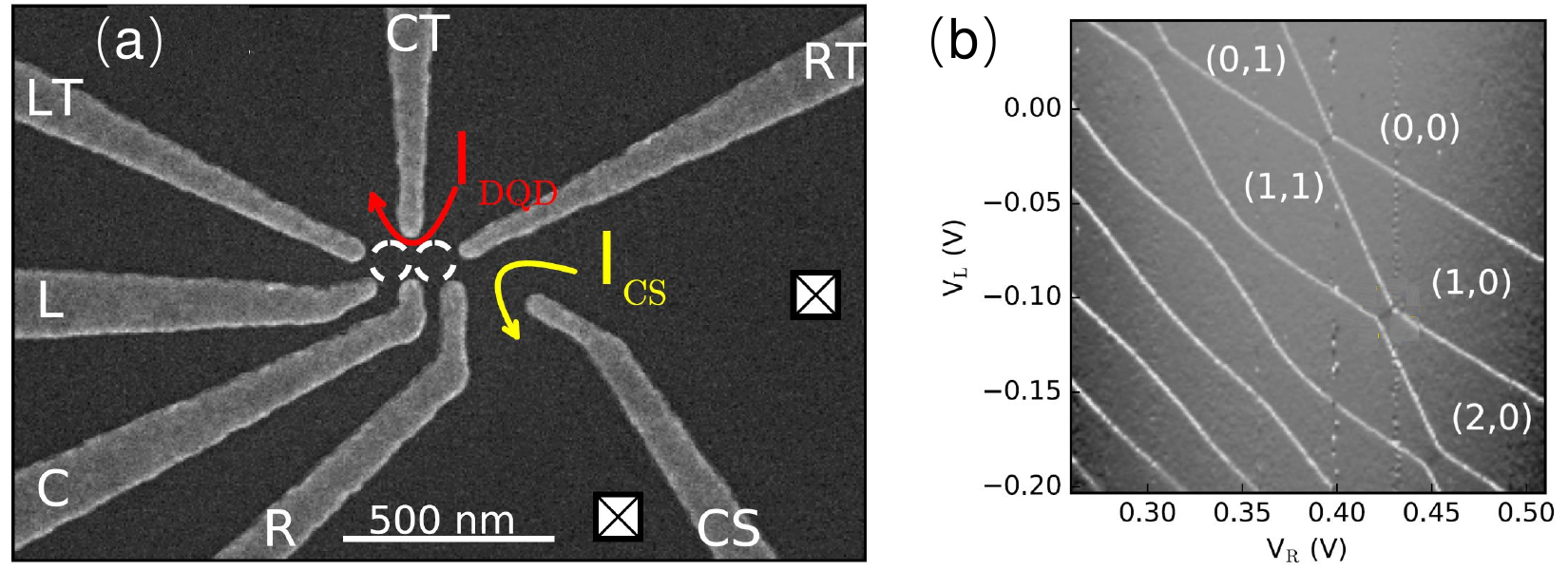}
\caption{(a) A double quantum dot based on a GaAs 2DHG. (b) Charging diagram, obtained from the current in the charge sensor, $I_{\rm CS}$, as function of voltages applied to gates L and R. Both panels are adapted from Ref.~\cite{Sachrajda17}.}
\label{GaAs_dQD}
\end{figure}

\emph{General device operation.} An example of a planar double quantum dot in GaAs is shown in Fig.~\ref{GaAs_dQD}(a). The holes are strongly confined along the $z$ direction (perpendicular to the 2DHG) thanks to the barrier induced by band offset at the GaAs/AlGaAs heterointerface, together with the voltage applied to a global top gate [not shown in Fig.~\ref{GaAs_dQD}(a)].
In addition, there is a smooth in-plane potential generated by voltages applied to the top gates labelled in Fig.~\ref{GaAs_dQD}(a). The 2DHG is locally depleted, to confine holes around minima of the potential in the $xy$-plane. The top gates enable the control of on-site energies of individual quantum dots, as well as tunneling barriers between them. It is usual to specify the charge state with the occupation number of each dot, e.g., $(n_{\rm L},n_{\rm R})$ for a double quantum dot (where L/R refers to the left/right dot). Figure~\ref{GaAs_dQD}(b) shows the stability diagram obtained by controlling the voltages applied to two of the gates. In this case, changes in the charge state are inferred from the current $I_{\rm CS}$ through a nearby quantum point contact, serving as a charge sensor. Besides charge sensing, the system can be probed by measuring the current $I_{\rm DOT}$ through the device [see Fig.~\ref{GaAs_dQD}(a)], under various operating conditions. This tunneling spectroscopy approach is very versatile. Application of a finite source-drain bias, or even oscillating voltages at the gates (e.g., inducing EDSR or photon-assisted tunneling \cite{Sachrajda18}), makes it possible for $I_{\rm DOT}$ to probe the excited states. The knowledge of spectral properties of the double quantum dot allows to extract various important parameters, such as tunneling amplitudes and the $g$-factors.

\emph{Spin-flip tunneling}. The (1,0) and (0,1) charge configurations, where only one hole is present, are already of considerable interest, due to a non-trivial interplay of Zeeman and tunneling interactions \cite{Sachrajda18,Sachrajda19,Sachrajda19relax,Sachrajda22}. The (1,0) configuration is realized at sufficiently large negative detuning $\epsilon= \epsilon_{\rm L}-\epsilon_{\rm R}$ (where $\epsilon_{\rm L/R}$ are on-site energies). In this case, the ground state at zero magnetic field is a Kramers doublet commonly denoted $\ket{L\Uparrow},\ket{L\Downarrow}$. Instead, a positive detuning gives the (0,1) single-hole states $\ket{R\Uparrow},\ket{R\Downarrow}$. Tunneling becomes important around $\epsilon \approx 0$, where the hole is in a coherent superposition of L/R states. In the device of Fig.~\ref{GaAs_dQD}(a), it was demonstrated that the tunnel amplitudes can be modulated through gate C in a large interval between $0.1-100~\mu$eV \cite{Studenikin21review}. Furthermore, in terms of the above basis states, tunneling is characterized by a `spin-conserving' ($t_{\rm N}$) and `spin-flip' ($t_{\rm F}$) amplitude. This should be expected in general since, as discussed at length,  $\ket{L\Uparrow},\ket{L\Downarrow}$, $\ket{R\Uparrow},\ket{R\Downarrow}$ are not pure spin states in the presence of strong spin-orbit interactions. Remarkably, $t_{\rm N}$ and $t_{\rm F}$ were found to be of similar magnitude. For example, for an applied magnetic field $B_z \simeq$1.4~T, a value $t_{\rm F} \simeq 0.2~\mu$eV, slightly larger than $t_{\rm N}\simeq 0.15~\mu$eV, was found in Ref.~\cite{Sachrajda18}. In the strong-tunneling regime, approximately equal values $t_{\rm N}\simeq t_{\rm F}\simeq 100~\mu$eV were reported around $B_z\simeq 1$~T \cite{Sachrajda19}, while $t_{\rm N} > t_{\rm F}$ for larger $B_z$. The relative strength of $t_{\rm N}$ and $t_{\rm F}$ has a significant dependence on $B_z$, as the two tunneling amplitudes are affected differently by orbital effects of an out-of-plane magnetic field \cite{Sachrajda18}. 

\emph{Pauli spin blockade.} Insight into the microscopic form of the spin-orbit coupling has been gained by studying the Pauli spin blockade \cite{Hamilton16}. Pauli spin blockade is a fundamental transport phenomenon for double quantum dots where, in the ideal case of conserved total spin, transitions from the $T(1,1)$ spin triplets to the (2,0) charge configurations are not allowed. Starting from the (1,1) charge state, the $T(1,1) \to T(2,0)$ process between triplets is normally not available at small detuning. In fact, $T(2,0)$ involves an excited orbital of the left dot, to satisfy the Pauli principle, thus has a high energy. Furthermore, the transition from triplet states to the singlet, $T(1,1) \to S(2,0)$, is forbidden when total spin is conserved. As a consequence, when the the double dot gets trapped in a $T(1,1)$ state, the current $I_{\rm DOT}$ from the right to the left reservoir is blocked \cite{Ono02}.

For holes, however, the large values of $t_{\rm F}$ imply a strong violation of the spin selection rule. Because spin-flip tunneling is not generally blocked, the study of the excited states though tunneling spectroscopy is facilitated by the existence of additional transport channels, which are normally suppressed \cite{Sachrajda17,Sachrajda21}. An important exception occurs when the external magnetic field $\vec{B}$ is along the effective spin-orbit coupling field $\vec{B}_{\rm SO}$ associated with the tunneling Hamiltonian~\cite{Nazarov09}. Then, the total spin along that direction is conserved, recovering the Pauli spin blockade. A study of the anisotropic dependence of the leakage current on $\vec{B}$ was performed in Ref.~\cite{Hamilton16}, where the direction of $\vec{B}_{\rm SO}$ was found to be consistent with a dominant Dresselhaus mechanism. As explained further below, this finding is in agreement with recent measurements of the relaxation time, $T_1$ \cite{Sachrajda19relax}. 

\emph{$g$-factors.} As we have discussed in Sec.~\ref{sec:SOC_qdots}, an important consequence of strong spin-orbit coupling is a significant dependence of hole $g$-factors on the confinement potential. For planar quantum dots, the tight confinement along $z$ results in hole carriers which have predominant heavy-hole character, i.e., atomic angular momentum $J_z \simeq \pm 3/2$ (for a more precise description, see Sec.~\ref{sec:SOC_general}). 
Because the heavy-hole states differ in angular momentum by $\Delta J_z \simeq 3$, the coupling between these states via an in-plane magnetic field (and the associated Zeeman interaction) is highly suppressed.
Accordingly, as in Eq.~(\ref{Hzeeman}), the in-plane heavy-hole effective $g$-factors are small. Indeed, a very small value $g_\parallel = -0.04 \pm 0.04$ was measured in Ref.~\cite{Sachrajda17}, similar to what is found in 2DHGs \cite{Marie99,Winkler00anisotropy,Korn10}. Instead, the out-of-plane heavy-hole effective $g$-factor of an (isolated) dot was measured in the range $g_z \simeq 1.1 - 1.4$ \cite{Sachrajda22}. 

Variations of $g_z$ are related to the influence of the confinement potential.
This influence can be leveraged to enable electrical control of the effective $g$-factors through gate voltages.
In Ref.~\cite{Sachrajda22}, it was shown that the value of $g_z$ depends on the voltage applied to the middle gate [labelled C in Fig.~\ref{GaAs_dQD} (a)].
It is instructive to look at the mechanism behind such variations, since the primary purpose of gate C is to modify the tunneling barrier between dots. Measurements, however, were performed in a regime of large detuning, where the modulation of tunneling amplitude has negligible effect on the hybridization of charge states. Therefore, the observation of a tunable value $g_z \simeq 1.1 - 1.4$ does not reflect a large variation of the tunneling amplitude, but is due to relatively minor changes in the confinement potential of a single dot. For the same reason, there is typically a $\sim 5\%-10\%$ difference in $g$-factors of the two (isolated) dots \cite{Sachrajda22}. More dramatic effects can be observed when the charge states are hybridized. A strongly reduced value $g_z=0.85$ was obtained in the regime of large tunnel couplings and zero detuning \cite{Sachrajda19}, where the Zeeman-split eigenstates are nontrivial mixtures of all single-hole states $\ket{L \Uparrow},\ket{L \Downarrow},\ket{R \Uparrow},\ket{R \Downarrow}$.  

\emph{Spin readout and relaxation time.} The $\ket{L \Uparrow} \to \ket{L \Downarrow}$ spin relaxation time has been measured in Ref.~ \cite{Sachrajda19relax} as function of $B_z$ using a  specialized spin-readout protocol. In general, spin readout in gated quantum dots is based on \emph{spin-to-charge conversion}, i.e., it relies on processes where different spin states lead to distinct charge states. An explicit example (based on Pauli spin blockade)  will be described in the next section. To access short relaxation times, however, it is important to overcome bandwidth limitations in charge sensing, and in Ref.~\cite{Sachrajda19relax} it was necessary to rely on a more sophisticated latched charge readout. In short (see Ref.~\cite{Sachrajda19relax} for details), the readout protocol transfers a hole in the $\ket{L \Uparrow}$ state to a metastable (0,1) charge configuration (with $\sim 50\%$ fidelity) via resonant spin-flip tunneling.
In contrast, holes in the $\ket{L \Downarrow}$ state are made to tunnel out of the left dot, leaving the system in the $(0,0)$ charge state. By measuring the final charge state, it is possible to infer the spin of the initial state.

Relaxation times range from $T_1 \simeq 0.3~\mu$s (at $B_z = 1.5$~T) to $T_1 \simeq 60~\mu$s (at $B_z = 0.5$~T) and follow an approximate $\propto B_z^{-5}$ dependence. This behavior supports a dominant Dresselhaus spin-orbit coupling \cite{Loss04,bulaev2005spin}, consistently with transport measurements in the Pauli spin blockade \cite{Hamilton16}. The measured values of $T_1$ are in good agreement with theoretical values of the Dresselhaus spin-orbit coupling \cite{Coish20}. We note that hole-spin relaxation times are much shorter than in electron-based GaAs quantum dots, where an extremely long $T_1 \simeq 60$~s could be demonstrated around 0.6 T \cite{Zumbuhl18relax}.

\emph{Spin manipulation}. In this system, a robust EDSR signal was detected with a continuous drive  \cite{Sachrajda19, Sachrajda22}. The measurements of $g_z$ were actually performed through EDSR. Unfortunately, controlled single-spin rotations have not yet been demonstrated. However, coherent oscillations between singlet and triplet states have been observed \cite{Studenikin21review}, demonstrating that the double quantum dot is suitable to realize a $S-T_-$ qubit \cite{Petta10,Burkard21_review}. More precisely, deep in the (1,1) region, the ground state at finite $B_z$ is a fully polarized triplet $T_-(1,1)$, i.e., the spin configuration is $\ket{\Downarrow\Downarrow}$ (minimizing the Zeeman energy). On the other hand, at sufficiently large negative detuning the ground state becomes $S(2,0)$ since, as discussed for the Pauli spin blockade, the triplet states $T(2,0)$ have higher energy. It follows that, when total spin is conserved, the $S$ and $T_-$ states must cross in energy as function of detuning. For hole systems, the relatively large value of $t_{\rm F}$ gives a robust $S-T_-$ anticrossing point, which can be exploited to induce Landau-Zener transitions. Starting from the singlet ground state and applying Gaussian-shaped detuning pulses, a Landau-Zener-St\"{u}ckelberg-Majorana interference pattern induced by the $S-T_-$ anticrossing could be obtained \cite{Studenikin21review}. 

\subsection{Planar Ge quantum dots}  

As for GaAs quantum dots, development of hole-spin qubits in planar Ge was made possible by the availability of high-quality 2DHGs. Doped quantum wells can reach high mobilities exceeding $10^6$~cm$^2$(Vs)$^{-1}$ \cite{Dobbie12,Failla16}, and the first demonstration of a gate-controlled quantum dot \cite{Veldhorst18} was based on an undoped device with 500,000~cm$^2$(Vs)$^{-1}$ mobility. Further advances in fabrication are reported in Refs.~\cite{Sammak19,Scappucci21,Scappucci22}. Remarkably, in the short time span from 2018 to 2021 experiments have advanced from a tunable single dot to a fully controllable four-qubit processor \cite{Veldhorst21,Veldhorst21_arXiv}. A reason behind this rapid progress is the small in-plane effective mass $\sim 0.05m_0$ \cite{Scappucci19,Culcer21} which facilitates fabrication (planar Ge quantum dots can be relatively large). 

\begin{figure}
\centering
\includegraphics[width=0.9\textwidth]{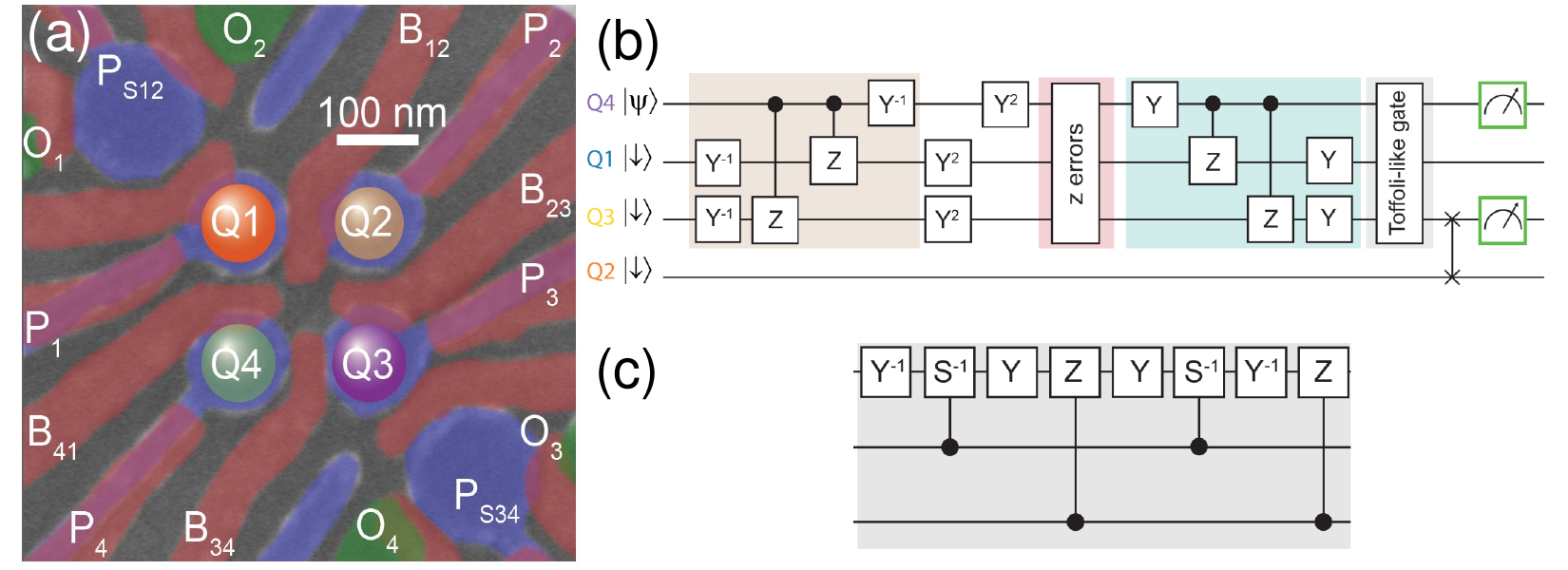}
\caption{(a) A 2D array of Ge quantum dots, realizing a four-qubit quantum processor. (b) Circuit diagram of a three-qubit phase flip code which was implemented in the device of panel (a). The code relies on Y gates (i.e., $\pi/2$ rotation around $y$) and a set of two-qubit gates summarized in Table~\ref{tab:2Qgates}. The beige box is the encoding step, performed with Y$^{-1}$ and CZ gates. Next, $\pi$ refocusing pulses (Y$^2$) are applied to the data qubits. The turquoise box is the decoding step. The error-correction step is implemented through a Toffoli-like gate, whose detailed circuit diagram is shown in panel (c) and involves two CS$^{-1}$ operations. The final gate of panel (b) is a SWAP between Q3 and the ancilla qubit Q2.  Panel (a) is adapted with permission from Ref.~\cite{Veldhorst21_arXiv}, while panels (b) and (c) are adapted with permission from Ref.~\cite{Veldhorst22}.}
\label{fig:Ge_planar}
\end{figure}

The four-qubit system is shown in Fig.~\ref{fig:Ge_planar}, together with a recently implemented algorithm \cite{Veldhorst22}. To operate the quadruple dot, it is helpful to define suitable linear combinations of the physically applied voltages which result in changes of a single on-site energy or tunnel coupling, without affecting other physical parameters of the qubits array (the so-called `virtual gates' \cite{Vandersypen17}). Efficient strategies to control a larger  number of qubits, such as a $4\times4$ array \cite{Borsoi22}, are being developed. In the following, we will describe in some detail how coherent manipulation of single-hole spins is realized in the quadrupole quantum dot, as well as the operation of singlet-triplet qubits in Ge double quantum dots.

\emph{Spin readout.} Single-shot spin readout in planar Ge quantum dots can be achieved through the same mechanism responsible for the Pauli spin blockade \cite{Veldhorst20_NC}. Consider a double quantum dot where, deep in the (1,1)  region, the low-energy states  are of the form $\ket{\Downarrow\Downarrow}$ and $\ket{\Downarrow\Uparrow}$  (due to the different $g$-factors of the two dots). $\ket{\Downarrow\Downarrow}$ is the $T_-(1,1)$ triplet which, by detuning the system to the (2,0) region, remains blocked in the same charge configuration (for a brief description of the Pauli spin blockade, see Sec.~\ref{sec:planar_GaAs}). On the other hand, $\ket{\Downarrow\Uparrow}$ has a finite overlap with the $S(1,1)$ singlet, and can evolve to the $S(2,0)$ ground state for an appropriate choice of the detuning ramp rate. This gives a spin-to-charge conversion process for the hole-spin in the right dot, $\ket{\Downarrow}\to (1,1)$ and $\ket{\Uparrow}\to (2,0)$, with the spin in the left dot acting as an ancilla qubit. As described for the GaAs planar dots, in changing the detuning from (1,1) to (2,0) a $S-T_-$ anticrossing point is  encountered. While this is a source of readout errors, the reduced visibility ($\nu \simeq 56\%$) was mainly attributed to triplet spin relaxation during charge sensing \cite{Veldhorst20_NC}, suggesting that a latched readout scheme could be advantageous \cite{Studenikin12,Coish18}. Indeed, later experiments adopted a latched Pauli spin blockade readout \cite{Veldhorst21}.

\emph{Single-spin manipulation and coherence.} In the same experiment  \cite{Veldhorst20_NC}, coherent single-spin manipulation was performed through EDSR, with Rabi frequencies $f_R = 57~$MHz and $93$~MHz for the two dots. Through a Ramsey sequence, a coherence time $T_2^* =380~$ns was extracted, likely limited by charge noise affecting the $g$-factor \cite{Veldhorst20_NC}. More generally, $T_2^*$ in various quantum dots was found in a range $\sim 150-800$~ns. The coherence time can be extended to $T^{\rm HE}_2 \sim 1-5~\mu$s with a Hahn echo sequence~\cite{Veldhorst20_NC,Veldhorst20,Veldhorst21} and up to $\sim 100~\mu$s by applying $N_\pi = 1500$ Carr-Purcell-Meiboom-Gill (CPMG) refocusing pulses \cite{Veldhorst21}. Spin lifetime measurements have obtained $T_1 = 1-16$~ms \cite{Veldhorst21}. The quality of the single-qubit gates is remarkable, as it was reported in the range $F=99.4 - 99.9\%$ for the four quantum dots of Ref.~\cite{Veldhorst21}. A more detailed analysis of single-qubit gates recently appeared \cite{Veldhorst21_arXiv}, showing that the strength of the EDSR drive can be optimized to almost reach the $F=99.99\%$ threshold (for one of the four qubits). While it is possible to obtain large values of the Rabi frequency, exceeding 100~MHz, faster spin manipulation comes at the price of unwanted errors and the optimal $F$ is realized at relatively small $f_R \simeq 10$~MHz. In the same array, the optimized fidelity of two other qubits exceeds $99.9\%$ and is close to $99.9\%$ for the last qubit. Interestingly, the fidelity shows little degradation when driving multiple qubits, indicating highly local cross-talk. A large value of $F\simeq 99.9\%$ is maintained upon simultaneous drive of any pair of neighboring qubits, and the fidelity is reduced to $ \simeq 99\%$ when all four qubits are driven simultaneously~\cite{Veldhorst21_arXiv}.

\emph{Two-qubit gates.}  Two-qubit gates can be realized via the exchange interaction. 
Because the exchange interaction is induced by a finite tunneling amplitude, only neighboring dots are coupled strongly enough to implement two-qubit gates effectively.  
Furthermore, the coupling strength, $J$, can be controlled through the tunnel barrier between the dots. Of special interest is the (1,1) configuration with equal on-site energies for the two quantum  dots. As first demonstrated for electron-spin qubits \cite{Hunter16,Kuemmeth16}, at this symmetric point the two-qubit gates are more resilient to charge noise, since the value of $J$ becomes insensitive (at leading order) to small variations of the detuning $\epsilon$. In demonstrating  a controllable exchange interaction in Ge planar dots, a tunable $J/h$ in the range $\sim 35- 65$~MHz was realized around the symmetric point \cite{Veldhorst20}. Tunability of the exchange coupling in the range $J/h \simeq 8 - 54$~MHz was reported by a more recent experiment~\cite{Veldhorst22rvb}. 

The discussion of various two-qubit gates can be simplified by neglecting the flip-flop terms, e.g., setting $J \boldsymbol{\sigma}_1 \cdot \boldsymbol{\sigma}_2 \approx J \sigma_{z,1} \sigma_{z,2}$, where $z$ is the direction of the magnetic field. This approximation becomes accurate when the difference in Zeeman energy between $\ket{\Uparrow \Downarrow}$ and $\ket{\Downarrow \Uparrow}$ is much larger than $J$. In practice, EDSR resonant frequencies of neighbouring dots can differ by a few hundred MHz (the difference is around 260 MHz in Ref.~\cite{Veldhorst20}), which is indeed larger than typical values of $J/h$. Note that, since the eigenstates of the Zeeman Hamiltonian are identified with the logical states $\ket{0},\ket{1}$ of the qubit, here the spin quantization axis is chosen conventionally along $z$ (even if the external magnetic field is physically applied in-plane).

For an interaction of the form  $J \sigma_{z,1} \sigma_{z,2}$, the EDSR frequency of qubit 1 depends on the state of qubit 2 (either $\ket{\Downarrow} \equiv \ket{0}$ or $\ket{\Uparrow} \equiv \ket{1}$). In the presence of an always-on exchange interaction, applying an EDSR drive  at one of the two available frequencies leads to a conditional rotation (CROT) of qubit 1, with qubit 2 acting as a control \cite{Veldhorst20,Veldhorst21}. A controlled-X (CX) gate is a conditional $\pi$-rotation, which could be realized with a gate time as fast as 55 ns \cite{Veldhorst20}. For the four-qubit quantum processor of  Ref.~\cite{Veldhorst21}, CX gates were executed on all four pairs of neighbouring qubits, with gate times under 100 ns. More advanced controlled rotations involve turning on two or three exchange couplings simultaneously (e.g., $J_{12}, J_{14} \neq 0$, where 1 is the target qubit while 2 and 4 are the control qubits). In this case, the EDSR resonance of the target qubit splits in four or eight lines, depending on the number of control qubits \cite{Veldhorst21}. 

\begin{table}
\centering
\begin{tabular}{l|l}
\hline\hline
  Gate & Protocol \\
  \hline
  CROT & EDSR with always-on exchange interaction \cite{Veldhorst20,Veldhorst21} \\   
	CPHASE & Pulsed exchange interaction $J(t)$ \cite{Veldhorst21} \\
	CX  & Conditional $\pi$-rotation (special case of CROT) \cite{Veldhorst20,Veldhorst21}  \\
  CZ & Based on CPHASE, with a $\pi$ phase \cite{Veldhorst21} \\
	CS$^{-1}$ & Based on CPHASE, with a $-\pi/2$ phase \cite{Veldhorst22}\\   
	SWAP & Oscillating $J(t)$, resonant to the $\ket{\Downarrow\Uparrow} -\ket{\Uparrow \Downarrow}$ transition \cite{Veldhorst22} \\
	$i$-Toffoli  & Similar to CROT, with $J_{12},J_{14}\neq 0$ (qubit 1 is the target) \cite{Veldhorst21} \\ 
	Toffoli-like & Combination of CX, CS$^{-1}$, and single-qubit gates \cite{Veldhorst22}\\
	\hline\hline
\end{tabular}
\caption{Summary of two- and three-qubit gates implemented in Ge planar quantum dots. A resonant four-qubit gate (extension of CROT) was also discussed in Ref.~\cite{Veldhorst21}.  }\label{tab:2Qgates} 
\end{table}

Besides controlled rotations, faster ($\sim 10$ ns) two-qubit gates were implemented through a pulsed exchange interaction. With an interaction $\propto \sigma_{z,1} \sigma_{z,2}$, a $J(t)$ pulse will induce a relative phase between parallel ($\ket{\Downarrow\Downarrow}$, $\ket{\Uparrow \Uparrow}$) and antiparallel ($\ket{\Downarrow\Uparrow}$, $\ket{\Uparrow \Downarrow}$)  states, allowing to realize a CPHASE gate. For a $\pi$ phase difference (and after correcting for single-qubit phases), a controlled-Z gate (CZ) could be implemented. All possible CZ gates between qubit pairs were demonstrated, with operation times under 10 ns  \cite{Veldhorst21}. In a similar way, the CPHASE gate leads to a controlled-S gate (CS), where the phase difference is $\pi/2$.
The recent demonstration of a phase flip code on three qubits required implementing a three-qubit Toffoli-like gate.
As shown in Fig.~\ref{fig:Ge_planar}(c), this three-qubit gate was constructed by  combining CZ, CS$^{-1}$, and single-particle gates \cite{Veldhorst22}. Finally, a SWAP gate exchanging antiparallel spin states, $\ket{\Downarrow\Uparrow} \leftrightarrow \ket{\Uparrow \Downarrow}$, was realized through an oscillating exchange pulse resonant with the  $\ket{\Downarrow\Uparrow} -\ket{\Uparrow \Downarrow}$ transition~\cite{Veldhorst22}. Implementation of this gate relies on the presence of flip-flop terms which couple the two antiparallel states. For convenience, we present a summary of all these two-qubit gates, together with the basic mechanism behind their realization, in Table~\ref{tab:2Qgates}.

\emph{Quantum algorithms.} Rather elaborate quantum algorithms could be implemented using the set of gates described above. In Ref.~\cite{Veldhorst21}, four qubits were entangled into a Greenberger-Horne-Zeilinger (GHZ) state and subsequently brought back to a product state. Although the fidelity for generating the GHZ state was not reported, parity readouts at intermediate steps and in the final state demonstrate that coherence is maintained throughout the whole protocol. Error correction codes were explored in Ref.~\cite{Veldhorst22}. The most sophisticated scheme (a phase flip code on three qubits) is reproduced in Fig.~\ref{fig:Ge_planar}(b). In both cases, an interesting point is that coherence could be improved by including refocusing $\pi$-pulses at selected steps of the time evolution, which demonstrates the benefits of combining quantum algorithms with dynamical decoupling. Furthermore, great flexibility in controlling the exchange interactions $J_{ij}$ for the four pairs of neighboring qubits has recently allowed the simulation of resonating valence bond (RVB) states of the $2\times2$ array~\cite{Veldhorst22rvb}. When operated under equal exchange couplings, $J_{12}=J_{34}=J_{14}=J_{23}$, the $s$-wave and $d$-wave symmetry RVB states are eigenstates of the system. 
Explicitly, these states can be written as $|s\rangle = (\ket{S_{12}}\ket{S_{34}}-\ket{S_{14}}\ket{S_{23}})/\sqrt{3}$ and $\ket{d}=\ket{S_{12}}\ket{S_{34}}+\ket{S_{14}}\ket{S_{23}}$, where $\ket{S_{ij}}$ is a singlet of spins $i$ and $j$. Preparation of these correlated spin states, as well as coherent oscillations in the two-dimensional subspace spanned by them, have been demonstrated~\cite{Veldhorst22rvb}.

\emph{Singlet-triplet qubits.} So far we have discussed qubits formed by individual spins.
However, it is also possible to define a qubit based on two suitable states of multiple coupled spins. In double quantum dots, one option is the $S-T_-$ qubit mentioned at the end of Sec.~\ref{sec:planar_GaAs}. Another attractive possibility is to consider the unpolarized spin states $(\ket{\Uparrow\Downarrow} \mp \ket{\Downarrow\Uparrow})/\sqrt{2}$, forming the so-called $S-T_0$ qubit \cite{Petta05}. To operate this two-level system, one can ramp the detuning rapidly from $S(2,0)$ to the (1,1) region, traversing the $S-T_-$ anticrossing diabatically, to reach $S(1,1)$. The singlet, however, is not an eigenstate of the system. In hole systems, if the difference in $g$-factors is large, the singlet can be significantly coupled to the unpolarized triplet $T_0(1,1)$ in the following way \cite{Katsaros21}: $H_{ST} = \frac12 \Delta g\mu_B B\left(\ketbra{S}{T_0}+\ketbra{T_0}{S}\right)-J(\epsilon)\ketbra{S}{S}$, where $B$ is the external magnetic field, $\Delta g$ is the difference in $g$-factors between the two quantum dots, and $J(\epsilon)$ is the exchange energy. Crucially, $J(\epsilon)$ strongly depends on  detuning. This dependence enables the control of the axis of rotation and the precession frequency of the effective qubit. Readout is performed by returning to the (2,0) region and relying on the Pauli spin blockade.

Coherent $S-T_0$ oscillations in a planar Ge double quantum dot can approach a frequency of 100~MHz at $B_z=3$~mT\cite{Katsaros21}. The dephasing time can reach  $T_2^* \simeq 1~\mu$s for relatively large detuning and a smaller $B_z = 0.5$~mT \cite{Katsaros21}. The latter regime mitigates the effect of fluctuations in $J$ and $\Delta g_z$, induced by charge noise. On the other hand, the best quality factor $Q=52$ was achieved at $B=3$~mT, when the oscillation frequency is larger. Coherent two-axis control of the $S-T_0$ qubit is possible and a CPMG decay time $T_2^{\rm CPMG}\simeq 4.5~\mu$s was obtained applying a sequence with $N_\pi =2$ $\pi$-pulses. The coherence time  can be further prolonged to $T_2^{\rm CPMG}\simeq 135~\mu$s, for $N_\pi =512$. Furthermore, the interplay between $S-T_0$ and $S-T_-$ coherent oscillations in different regimes was analyzed by Ref.~\cite{Katsaros22}. That study of coherent singlet-triplet oscillation also led to a full characterization of the device: Site dependent and anisotropic $g$-factors were found, with values $g_z \simeq 7 $ and $g_x \simeq 0.26 $ for the left dot and $g_z \simeq 5$ and $g_x\simeq -0.16$ for the right dot \cite{Katsaros22}. As previously shown in Ref.~\cite{Katsaros21}, the difference in $g_z$-factors can be modulated through the center gate, from $\Delta g_z\sim 1.5$ to above $2.2$.  The detailed form of the spin-flip tunneling term was determined as  $t_x \sigma_x + t_y \sigma_y$, with $t_x \simeq 0.13~\mu$eV and $t_y\simeq -0.37~\mu$eV  \cite{Katsaros22}. These values give a total strength $t_{\rm F} =\sqrt{t_x^2+t_y^2} \simeq 0.39~\mu$eV (while the spin-conserving tunneling amplitude is $t_{\rm N}\simeq 11~\mu$eV). More recently, coherent $S-T_-$ oscillations were also demonstrated in the four-qubit array of Ref.~\cite{Veldhorst22rvb} for all four pairs of neighboring quantum dots. At an external field of 1 mT, the oscillation frequency is between 1~MHz and 2.6 MHz, depending on the pair, and typical dephasing times of several $\mu$s were obtained (the longest dephasing time, obtained for one of the four pairs, is  $11~\mu$s \cite{Veldhorst22rvb}).

\subsection{Hole-spin qubits in nanowires} \label{sec:nanowires}

\begin{figure}
\centering
\includegraphics[width=1\textwidth]{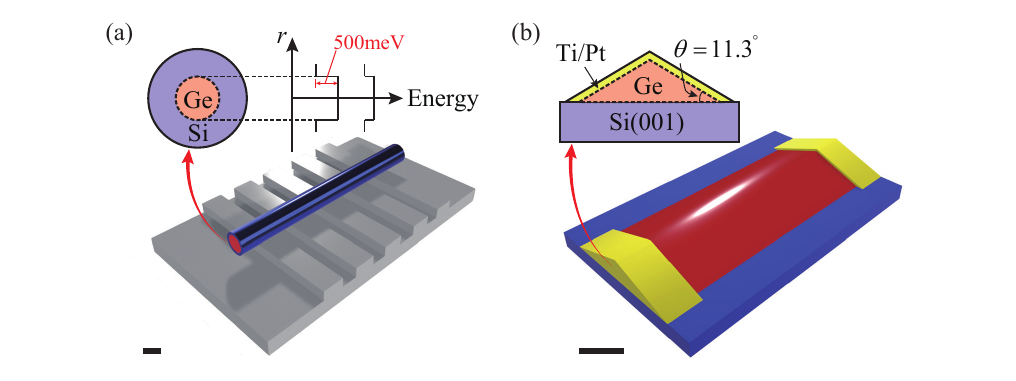}
\caption{(a) Ge/Si core/shell nanowire on top of metallic gates. The enlarged plot depicts the cross-section of the nanowire, where the valence and the conduction bands are shown schematically as a function of radial distance $r$. (b) Typical configuration of a Ge hut wire, on top of its silicon substrate. Scale bar: Rough indication of a $10\mathrm{nm}$ length.}
\label{NW_dot}
\end{figure}

Quantum dots formed in nanowires are another attractive option to develop spin qubits. An early demonstration of EDSR with hole spins was based on an InSb nanowire, where both single and double quantum dots could be realized \cite{Kouwenhoven13}. The confinement potential localizing the holes to the quantum dots is generated by applying voltages to thin metallic gates fabricated in close proximity to or on top of the nanowire (see Fig.~\ref{NW_dot} for a representation of two example configurations). In Ref.~\cite{Kouwenhoven13}, evidence of depleting the quantum dots to the last hole is  provided. This evidence is obtained by transport measurements since charge sensing is more challenging in nanowires than lateral quantum dots (a charge sensor should be fabricated as an external element \cite{Marcus07}). A strongly anisotropic $g$-factor  was measured, in the range $g \simeq 0.5 - 4$, and the Pauli spin blockade in the double quantum dot could be studied. This same 2013 experiment demonstrated EDSR through a continuous drive, but coherent manipulation of the hole spin was not yet realized.

For hole spins in nanowires, coherent Rabi oscillations were reported in 2018 based on a Ge `hut wire'~\cite{Katsaros18a}. Recent progress has focused on group IV elements, rather than III-V nanowires, and two main types are illustrated in Fig.~\ref{NW_dot}: Ge/Si core/shell nanowires (left) and the aforementioned Ge hut wires (right). The presence of strong spin-orbit interaction is a common feature which, recently, has allowed to reach ultrafast single-spin manipulation, with Rabi frequencies surpassing 400~MHz for both platforms \cite{Froning2021,Guo22a}. The presence of a strong spin-orbit interaction should also facilitate the coupling of nanowires to superconducting transmission lines \cite{Loss08nanowire,Loss13qed,Loss22}. As discussed below, there is some progress in this direction with Ge nanowires. Most notably, however, strong spin-photon coupling has been reported by a recent experiment based on a Si nanowire transistor \cite{Franceschi22a}. More detail will be given in the next subsection (on Si devices).

\emph{Ge/Si core/shell nanowires.} These nanowires, schematically illustrated in Fig.~\ref{NW_dot}(a), have a cylindrical cross-section with a Ge core, of typical diameter $\sim 10$~nm, and a Si shell providing strong confinement due to the large offset between the valence bands of silicon and germanium. Details on fabrication can be found in Refs.~ \cite{Lieber98,Duan2000,Lieber02}. The Fermi energy lies entirely within the germanium valence band and free holes are confined to the Ge core~\cite{Walle86,Lieber05}. Transport measurements reflect the high quality of the interface between the undoped Ge/Si materials, allowing to observe ballistic transport up to room temperature and a mobility of 600~cm$^2$(Vs)$^{-1}$~\cite{Lieber05,Lieber06,Guo10}. The strong spin-orbit interaction in these nanowires can be characterized through the spin-orbit length $\lambda_{\rm so}$ (see at the end of Sec.~\ref{sec:SOC_general}), which was found between $20 - 65$~nm in various devices \cite{Guo10,Lieber14,Froning2021a}, and could be tuned by a factor of 5 using a back gate \cite{Guo10}. In a recent experiment, the tunable value of $\lambda_{\rm so}$ was estimated (assuming a heavy-hole effective mass) to range from 28 nm to an extremely short value of 3 nm~\cite{Froning2021}.

Quantum dots can be formed through electrostatic potentials from metallic gates next to the core/shell nanowire \cite{Marcus07,Lieber08,Froning2018}. However, it has proved difficult to reach the single-hole regime in these devices and typical occupations are of the order of tens of holes. For example, the working point chosen in Ref.~\cite{Marcus14} is around the charge state $\sim(70,10)$ and about 15 holes were confined in the dots of Ref.~\cite{Froning2021a}. This does not prevent experiments analogous to the ones in the single-hole regime. If the charge state of a double quantum dot $(n_{\rm L},n_{\rm R})$ is such that $n_{\rm L/R}$ give `closed-shell' configurations, then $(n_{\rm L}+1,n_{\rm R}+1)$ behaves effectively as (1,1), i.e. as a system with a single hole-spin trapped in each quantum dot \cite{Marcus14}. 

Early studies of spin coherence were based on detuning pulses~\cite{Marcus12,Marcus14}. Like in lateral quantum dots, initialization and readout can be achieved through the Pauli spin blockade \cite{Marcus12}. Using this scheme, spin-lifetime measurements are possible. Values in the range $T_1 \simeq 0.2 - 0.6$~ms (depending on the strength of the magnetic field) were found \cite{Marcus12}. However, a much shorter $T_1 \simeq 0.2 - 0.8~\mu$s was later reported in Ref.~\cite{Marcus14}, possibly due to a strong dependence of $T_1$ on the device and/or detuning.  In the same experiment, a dephasing time $T_2^* \simeq 180$~ns was extracted. The exponential (rather than Gaussian) decay of coherence suggests a negligible influence of nuclear spins \cite{Marcus14}.

Coherent manipulation through EDSR was recently achieved~\cite{Froning2021}. Rabi oscillations reaching $f_R \simeq 250$~MHz (at the largest applied drive) and two-axis control were demonstrated. The dephasing time $T_2^* \simeq 11$~ns extracted from a Ramsey sequence is much shorter than previous reports, but can be extended to $T^{\rm HE}_2 \simeq 250$~ns using Hahn echo. While these figures of merit are obtained at a specific working point, Ref.~\cite{Froning2021} also finds that $f_R$ and the decay time $T_2^{\rm Rabi}$ of the Rabi oscillations depend sensitively on the voltage, $V_M$, applied at one of the gates. For a small change $\Delta V_M \simeq 30$~mV, the Rabi frequency could be increased from $f_R =31$ to 219~MHz, which is accompanied by a reduction of $T_2^{\rm Rabi}$ from $59$ to 7~ns. The increase of $f_R$ and the reduction of $T_2^{\rm Rabi}$ are attributed to the strong effect of $V_M$ on the spin-orbit coupling in the nanowire~\cite{Kloeffel_PRB2011}. By optimizing the values of the applied gate voltages and for a large drive amplitude, a maximum Rabi frequency $f_R \simeq 435$~MHz could be reached~\cite{Froning2021}. 

In these nanowires, initial steps towards coupling hole-spin qubits to microwave photons were taken. A coherent charge coupling of 55~MHz between a double quantum dot (acting as charge qubit) and a superconducting transmission line resonator was realized~\cite{Wang2019}.

\emph{Ge hut-wires.} Germanium hut wires, discovered in the early 1990s \cite{Lagally90}, are self-assembled structures obtained by growing germanium on a silicon substrate. Due to the $4\%$ lattice mismatch between the germanium and silicon crystals, first pyramid-shaped and later hut-shaped germanium islands are formed on top of the Si(001) substrate, depending on the duration of the growth\cite{Tersoff93,Drucker08}. By using a catalyst-free method, high quality Ge hut wires with length-width ratio exceeding $1000$ could be fabricated \cite{Schmidt12}. The geometry of such nanowires is schematically shown Fig.~\ref{NW_dot}(b). They have typical height and width of $2$~nm and $18$~nm \cite{Schmidt12,Katsaros17}, respectively. To minimize  the surface energy, the slope between the side surfaces of the hut and the substrate plane is fixed at $11.3^\circ$, corresponding to $\{105\}$ facets  \cite{Tersoff93,Schaffler09,Schmidt12,Bauer04}.  Hut wires with similar structure have also been found for $\mathrm{Si}_{1-x}\mathrm{Ge}_x$ instead of Ge \cite{Schaffler14}.

Like in core/shell nanowires, quantum dots based on hut wires are currently operated in the multi-hole regime. Here however, due to the relatively `flat' geometry, the holes can have a predominant heavy-hole character, reflected by the strong anisotropy of the $g$-factors. Typical values are $g_z \simeq 2 - 4.5$ for the direction perpendicular to the substrate, and $g_{x,y} \lesssim 0.5$ for the in-plane direction \cite{Katsaros2016,Katsaros18a,Guo21,Guo22b}. The precise values depend on the device, as well as the quantum dot occupation. A maximum ratio $g_z/g_y \simeq 18 $ was observed inside a certain Coulomb diamond but the anisotropy tends to decrease for Coulomb diamonds with more holes, presumably because of a larger admixture with light-holes \cite{Katsaros2016}.

Insight into the spin-orbit coupling has been gained from measurements of the leakage current in the Pauli spin blockade regime \cite{Guo22a,Guo22b}. In one experiment the spin-preserving tunnel coupling was found to be in a  range from $t_{\rm N} \simeq 0.4~\mu$eV to $14.5~\mu$eV, while the spin-flip tunneling amplitude remains relatively constant at a larger value $t_{\rm F} \sim 20~\mu$eV. This translates to a tunable spin-orbit length in the wide range $\lambda_{\rm so} \sim 2 - 48.9$~nm \cite{Guo22a}. The shortest value, $\lambda_{\rm so} \sim 2$~nm, is compatible with EDSR measurements, where a spin-orbit length as short as $\lambda_{\rm so} \simeq 1.4$~nm was estimated from the dependence of the Rabi frequency on drive amplitude \cite{Guo22a}. By studying the anisotropic dependence of the leakage current on the applied magnetic field, an in-plane spin-orbit field forming an angle of $31 ^\circ$ with the nanowire was found~\cite{Guo21}. The nontrivial direction indicates a significant contribution from Dresselhaus spin-orbit coupling, which may be induced by interface inversion asymmetry.

Readout of the hole spin in a double quantum dot can be realized, as described  before, through the Pauli spin blockade \cite{Katsaros18a}. Alternatively, relying on two neighboring wires in a `T' configuration \cite{Katsaros17}, the same type of readout as Elzerman et al.~\cite{Elzerman04}, was demonstrated. In the latter scheme, the alignment of the chemical potential of an external reservoir with a single quantum dot is such that the upper (lower) energy state of the Zeeman-split doublet can (cannot) tunnel out into the lead. In Ref.~\cite{Katsaros18b}, one of the nanowires of the T configuration is both capacitively and tunnel coupled to a single quantum dot (realized at the end of the other wire). Accordingly, this wire can act as both a reservoir and a charge sensor, detecting tunneling events associated with the higher-energy $\ket{\Uparrow}$ state. Using such readout technique, spin lifetimes increasing with an approximate $B_z^{-3/2}$ dependence were measured. They range from $T_1 \simeq 15~\mu$s (at $B_z =1.5$~T) to  $90~\mu$s (at $B_z=0.5$~T)~\cite{Katsaros18b}.

Coherent control of the hole-spin qubit is possible through EDSR. A first demonstration could achieve a Rabi frequency of 140~MHz and, through a Ramsey sequence, a dephasing time  $T_2^* \simeq 130$~ns was measured \cite{Katsaros18a}. More recently, Rabi frequencies as large as $f_R = 542$~MHz were attained for EDSR signals mediated by the strong spin-orbit coupling (spin-orbit length $\lambda_{\rm so}\sim 1-2$~nm) mentioned above \cite{Guo22a}. In this case, values of the coherence time were found in the range $T^*_2 \simeq 42 - 84$~ns, depending on the applied power (to implement the $\pi/2$ pulses). With a Hahn echo sequence, the coherence time could be prolonged from  $T^*_2 \simeq 65$ ns to  $T^{\rm HE}_2 \simeq 523$~ns.

A potential advantage of hut wires over core/shell nanowires is that they directly grow on the two-dimensional substrate, facilitating the fabrication of more complex arrangements of multiple wires. Besides the spontaneously formed `T' configurations exploited for charge sensing, a method to fabricate structured arrays in a controlled manner, where the hut wires are formed along patterned trenches, is of particular interest \cite{Katsaros20}. Scalability could also be facilitated by coupling such spin qubits to superconducting transmission lines. Initial experiments demonstrated charge coupling of single \cite{Guo18} and double quantum dots \cite{Guo20} to microwave resonators, with a coupling strength $g_c/2\pi =148$~MHz ($15$~MHz) in the former (latter) case. Estimates based on the charge coupling still give relatively small values for the spin-photon coupling, around $2-4$~MHz \cite{Guo18,Guo20}. 

\subsection{Hole-spin qubits in silicon devices}

An attractive feature of hole-spin qubits hosted in silicon is that they can be fabricated using the complementary metal-oxide semiconductor  (CMOS) technology, making them compatible with industrial standards of nanofabrication. In order to perform quantum computation aimed at realistic applications, arrays of millions of qubits working together will likely be required. To tackle this demanding up-scaling task, semiconductor quantum dots based on well-established industrial fabrication processes could present crucial advantages, e.g., in terms of device uniformity and integration with classical circuitry \cite{Dzurak21}. 

\begin{figure}
\centering
\includegraphics[width=1\textwidth]{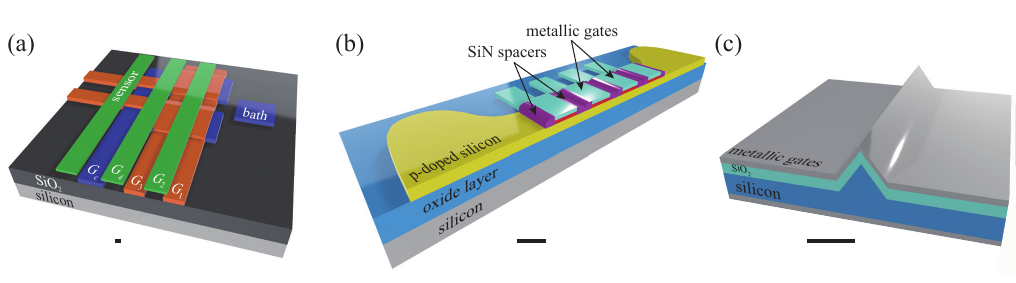}
\caption{Schematic plot for three types of silicon-based hole spin quantum devices. In each panel, the scale-bar corresponds roughly to a $ 10~\mathrm{nm}$ length. (a) Planar metal-oxide-semiconductor device, with top gates based on a multilayer Al-Al$_2$O$_3$ stack technology \cite{Hamilton13}. (b) Quantum dots based on a silicon nanowire field-effect transistor (in yellow). The nanowire channel has a typical width and height of $20$~nm and $10$~nm, respectively, and is partially $p$-doped in the source and drain regions. (c) Vertical structure of a silicon FinFET \cite{Zumbuhl18networks,Zumbuhl22}. The silicon fin is $ \sim 10~\mathrm{nm}$ wide and is covered by a $\sim 4.5\mathrm{nm}$ oxide layer, on top of which metallic gates are deposited (here the metallic gate is not on scale).}
\label{silicon_dot}
\end{figure}

Three main types of devices are illustrated in Fig.~\ref{silicon_dot}. In all of them, the quantum dots are formed by charge accumulation at the Si/oxide interface, induced by top metallic gates. Panel (a) represents a planar geometry \cite{Hamilton15,Hamilton18,Hamilton21}, while in panel (b) the quantum dots are based on a nanowire field-effect transistor (FET) \cite{Franceschi16a,Franceschi16b,Franceschi16c,Franceschi18,Franceschi21, Franceschi22a,Franceschi22b}. In the FinFET geometry of panel (c), similar to a nanowire, holes are confined at the top of a Si extrusion with triangular cross-section~\cite{Zumbuhl18,Zumbuhl18networks,Zumbuhl22}. Carriers are usually from p-doped regions near the quantum dots. However, in undoped FinFETs the carriers are from metallic NiSi electrodes, which also allow ambipolar operation of the device \cite{Zumbuhl18}. As usual, the top gates allow to control crucial parameters of the quantum dots, such as tunnel barriers between them and to the reservoirs, as well as the dot occupation. Single-hole occupation has been reached in quantum dots of type (a) \cite{Hamilton18,Hamilton21} and (b) \cite{Franceschi16c}. The FinFET device of Ref.~\cite{Zumbuhl22} also likely operates in the single-hole regime.

\emph{Spin-orbit interaction.} Similarly to quantum dots discussed in previous sections, the properties of hole-spin qubits in Si reflect the presence of a strong spin-orbit interaction, which can be studied through the leakage current in the Pauli spin-blockade regime \cite{Hamilton15,Franceschi16a,Zumbuhl21}, as well as through spectroscopic studies~\cite{Franceschi21,Franceschi22a} (where the Si quantum dots are coupled to resonators). A spin-orbit length of $\lambda_{\rm so} \sim 110$~nm was estimated in the planar double dot of Ref.~\cite{Hamilton15}, and $\lambda_{\rm so} \simeq 48$~nm in the FinFET of Ref.~\cite{Zumbuhl21}. In the nanowire device of Ref.~\cite{Franceschi21}, a spin-orbit gap of  $19~\mu$eV was measured at the $S-T_-$ anticrossing, which reflects the strength of the spin-flip tunneling amplitude $t_{\rm F}$. This energy scale is similar to the spin-conserving tunnel amplitude ($t_{\rm N}, t_{\rm F} \sim 10~\mu$eV in that experiment) .

A detailed analysis of the spin-flip tunneling amplitude was performed recently in the nanowire-FET double quantum dot of Ref.~\cite{Franceschi22a}, where $t_{\rm F}$ is a crucial parameter to achieve strong coupling of a single-hole spin (hosted in a double quantum dot) to a superconducting resonator. The strength of the spin-flip tunneling amplitude depends on the relative angle between $\hat{n}_{\rm SO}$ (the direction of the spin-orbit field) and $\hat{n}=\overleftrightarrow{g}\cdot {\bf B}$ (where $\overleftrightarrow{g}$ is the $g$-tensor, assumed equal for the two quantum dots: $\hat{n}$ is a natural quantization direction for the hole-spins). Therefore, $t_{\rm F}$  has a strongly anisotropic dependence on the magnetic field. In Ref.~\cite{Franceschi22a}, where the total tunneling amplitude is $t_c =\sqrt{t_{\rm N}^2+t_{\rm F}^2} \simeq  40~\mu$eV, the spin-flip amplitude is very small when $\hat{n}$ and $\hat{n}_{\rm SO} $ are almost parallel, giving $t_{\rm N} \simeq t_c \gg t_{\rm F}$. Instead, $t_{\rm F}$  approaches its maximum  value when $\hat{n}$ and $\hat{n}_{\rm SO} $ are nearly perpendicular. The latter regime gives $t_{\rm F} \simeq t_{c} \gg t_{\rm N}$ in this particular experiment, suggesting that the spin-orbit length is close to the distance between the two dots ($\lambda_{\rm so}\sim 80$ nm).

\emph{g-tensor.} The  effective $g$-tensor in these Si quantum dots is anisotropic, although not as strongly as for other types of hole-spin qubits, and can be significantly affected by small variations of the gate voltages \cite{Franceschi16c,Franceschi18,Hamilton21,Franceschi22b}. For example, in a quantum dot based on a  nanowire FET, it was possible to modulate the $g$-factor for the in-plane direction perpendicular to the channel from $g=1.85$ to 2.6 \cite{Franceschi16c}. A similar degree of tunability was demonstrated in the planar geometry~\cite{Hamilton21}. In general, the principal axes of the $g$-tensor are not aligned with the  crystallographic directions and do not have a simple correspondence with the geometry of the device. This is true for nanowire FET devices~\cite{Franceschi18,Franceschi22b} but also, more surprisingly, for planar quantum dots~\cite{Hamilton21}. In the latter case, despite the expectation of a predominant heavy-hole character, the largest $g$-factor is obtained along a direction forming an angle $\phi_0$ from the $z$ axis. Considering two distinct gate voltages, ($g_{\rm max}= 3.9$, $\phi_0=42^\circ$)  and ($g_{\rm max}= 1.7$, $\phi_0=70^\circ$) were measured, showing that not only the value of the $g$-factor, but also the direction of the principal axes is electrically tunable. The occurrence of large tilt angles was attributed to nonuniform stain induced by the metallic electrodes, which can have a strong influence on the light-hole/heavy-hole mixing~\cite{Hamilton21}.  As a reference for the typical anisotropy obtained in these devices, at the same two working points, Ref.~\cite{Hamilton21} finds a minimum $g$-factor of $1.4$ (giving $g_{\rm max}/g_{\rm min} \simeq 2.8$) and $0.3$ (giving $g_{\rm max}/g_{\rm min} \simeq 5.7$).

\emph{Hole-spin manipulation and coherence.} Coherent spin manipulation through EDSR  was demonstrated in 2016 with a nanowire FET device \cite{Franceschi16b}. Based on a double quantum dot with a $\sim(10,30)$ occupation, Rabi oscillations with a maximum frequency $f_R = 85$ MHz were realized. A dephasing time of $T_2^* \simeq 60$~ns was  obtained from the Ramsey fringes, while the Hahn echo amplitude decays with $T_2^{\rm HE} \simeq 245$~ns. These coherence times could be significantly improved in a recent experiment based on the same type of quantum dots, but where optimal operation points minimizing the effect of charge noise were identified \cite{Franceschi22b} (see below for more details).

Another notable development is the recent demonstration of coherent spin manipulation in a FinFET device \cite{Zumbuhl22}, performed in a large-temperature regime from 1.5~K to 4.2~K. Operating the qubits above 1~K  would enable a higher cooling power, which is an obvious advantage for designing large-scale devices with tightly integrated quantum hardware and classical control electronics. Rabi oscillations with frequency up to $f_R = 147$~MHz were demonstrated in this device and, from the Ramsey fringes, a coherence time of $T_2^* \simeq 200$~ns was obtained at 1.5~K for one of the two qubits. The coherence time is reduced to $T_2^* \simeq 90$~ns at  4.2~K. Correspondingly, through randomized benchmarking, a single-qubit gate fidelity $F=98.9\%$ was found at 1.5~K, which decreases to $F=97.9\%$ at 4.2~K. The dephasing time depends on the magnetic field, and a maximum value $T_2^* \simeq 440$~ns could be obtained, comparable to the dephasing timescale induced by nuclear spins. Furthermore, dynamical decoupling allows to reach $T_2^{\rm CPMG} \simeq 5.4~\mu$s, by applying a CPMG sequence with $N_\pi =32$.

\emph{Sweet spots.} As the $g$-factor of hole-spin qubits is sensitive to small fluctuations in the voltage $V$ of a nearby gate, identifying sweet spots at which $dg/dV=0$ is highly valuable. Experimentally, measurements of the voltage-dependent $g$-factor could provide evidence for such a stationary point in a planar CMOS quantum dot~\cite{Hamilton21}. More recently, the impact of sweet spots on the coherence of the hole-spin qubit was investigated for a nanowire-FET device \cite{Franceschi22b}. In this experiment, the direction of the external magnetic field was varied and two orientations were identified at which the Larmor frequency of the qubit becomes insensitive (to linear order) to fluctuations in the accumulation gate voltage. Due to the influence of a second more distant gate, the optimal orientation slightly deviates from one of the two sweet spots. The improvement in dephasing time (extracted from the Ramsey oscillations) is from $T_2^* \simeq 3~\mu$s (for a magnetic field along the nanowire FET)  to $T_2^* \simeq 6~\mu$s (at the optimal point). By applying a Hahn echo sequence, an even more significant enhancement, from $T_2^{\rm HE} \simeq 15~\mu$s  to $T_2^{\rm HE} \simeq 85~\mu$s, could be demonstrated. The longest coherence time is achieved with a CPMG sequence of $N_\pi = 256$ $\pi$-pulses, yielding $T_2^{\rm CPMG} \simeq 0.4$~ms \cite{Franceschi22b}. 

\emph{Strong spin-photon coupling.} A large spin-photon coupling was recently reported for a hole-spin qubit in a nanowire FET coupled to a superconducting resonator \cite{Franceschi22a}. Similarly to experiments with electron-spin qubits \cite{Mi18,Samkharadze18}, the coupling is based on the large dipole moment associated with the bonding/antibonding states of a single hole in a double quantum dot. In the absence of an applied magnetic field, a strong charge-photon coupling of $513~\mathrm{MHz}$ was deduced from the dispersive shift of the cavity frequency induced by the double quantum dot around zero detuning. At finite magnetic field, the spin-photon interaction depends crucially on the strength of the spin-flip tunneling amplitude $t_{\rm F}$. Consequently, like $t_{\rm F}$, the spin-photon interaction has a strongly anisotropic dependence on $\bf B$ (see discussion above). A maximal spin-photon coupling  of $g_s/2\pi=330~\mathrm{MHz}$ was achieved, which is much larger than the decay rate of the cavity ($\kappa/2\pi= 14$~MHz) and the spin decoherence rate ($\gamma_s/2\pi =2.5-17$~MHz). The cooperativity can be large as well: $4g_s^2/\kappa\gamma_s  = 1600$ for a magnetic field parallel to the nanowire (where $g_s$ is not maximal). Furthermore, a significant $\sim 1$~MHz spin-photon coupling could be demonstrated in the single-dot limit of a large detuning \cite{Franceschi16a}.

\emph{Acceptor dopants.} Finally, we address boron acceptor dopants in silicon, which represent another option for hole-spin qubits. Such dopants have been investigated in natural \cite{Simmons15,Simmons16} and isotopically purified silicon \cite{Kobayashi2021}, as well as gated structures \cite{Hamilton14,Simmons18,Yang22}. Characterizations of the electronic states of a single dopant, or two coupled acceptors \cite{Simmons16,Simmons18}, can be obtained from scanning tunneling spectroscopy \cite{Simmons13,Simmons15, Simmons16}  or transport measurements  \cite{Hamilton14,Simmons18}. Coherent spin manipulation was performed with pulsed electron spin resonance in an isotopically purified system \cite{Kobayashi2021}, showing a long CPMG coherence time of $9.2~\mathrm{ms}$. For such donors, the existence of sweet spots protecting the qubit from charge noise \cite{Salfi16}, as well as spin-echo envelope modulations as a sensitive probe of the anisotropic hyperfine interaction \cite{Philippopoulos19}, were recently discussed theoretically.

\section{Conclusion and outlook}

The rich physics of holes brings substantial opportunities to the development of spin qubits. Thanks to large spin-orbit couplings, the effective spin interactions are strongly affected by device geometry and electrostatic confinement. This dependence leads to great flexibility in engineering the properties of hole-spin qubits and enables local electric control of the hole-spin qubits. Concretely, fast single-spin manipulation becomes possible without the need for large external elements, such as striplines or micromagnets. These features, together with the absence of valley degeneracy and the small effective mass of Ge, are advantageous for the fabrication of large arrays of quantum dots with uniform properties, thus they will facilitate scaling up the current few-qubit experiments.

Due to the variety of available hole-spin qubits, the direction of future progress depends mainly on the specific implementation. A comparison between different  platforms might suggest the most natural and desirable developments for each type of qubit. For example, for lateral quantum dots in Ge it would be important to realize a few-qubit quantum processor which operates within the error-correcting thresholds (as recently achieved with electron-spin qubits \cite{Xue22,Noiri22,Mills22}). Spin-orbit coupling plays a relatively minor role in entangling gates, which rely on the exchange interaction. So, optimizing the two-qubit gates to reach a high fidelity, comparable to electron-spin qubits, might be within reach. Such improvements in the control of two-qubit operations would be also interesting for exchange-only architectures \cite{DiVincenzo00,Medford13,Burkard21_review}. 

Similarly, the recent demonstration of strong coupling between a Si hole-spin qubit and a superconducting resonator suggests that analogous experiments could be performed with other types of hole-spin qubits in the near future. Realizing a distant interaction between two hole spins coupled to a superconducting transmission line would be another important step towards more complex networks of interacting qubits. 

The influence of charge noise (mediated by the strong spin-orbit interaction) is a main obstacle to highly coherent spin manipulation of hole-spin qubits. Ideas about geometries mitigating this problem \cite{Salfi16,Culcer21npj,Loss21a,Veldhorst22theory} or realizing tunable spin-orbit interactions with large on/off ratios \cite{Kloeffel_PRB2011,KloeffelPRB2018,Mutter21,Loss21b} have now seen their first implementations \cite{Froning2021,Franceschi22b}. Further theoretical and experimental progress in this direction should be expected.

The rapid advances witnessed in recent years by hole-spin qubits will likely continue in the near future, establishing this system firmly as one of the most promising types of qubits. Since all the fundamental elements of quantum-information processing have been demonstrated in few-qubit devices, both electron-spin and hole-spin qubits might have a viable path towards large-scale architectures in front of them. However, the engineering challenges are still formidable and it is unclear which type of spin qubit will travel farther along the way. Adding to the uncertainty, long-term developments of spin-qubits have been characterized by shifts in the preferred material platform. Even the rapid rise of hole-spin qubits has, at its foundation, progress in fabrication. Crucial future advances could come from improved materials and nanostructure designs allowing, e.g., for reduced charge noise in certain devices. Such important developments, while difficult to anticipate, would further facilitate the field to reach its ambitious goals.  

\section*{Acknowledgments}

YF acknowledges support from NSFC (Grant No.
12005011). DC is supported by the Australian Research Council Centre of Excellence in Future Low-Energy Electronics Technologies (project number CE170100039). WAC acknowledges funding from the Natural Sciences and Engineering Research Council (NSERC) and from the Fonds de Recherche---Nature et Technologies (FRQNT). SC acknowledges support from the Innovation Program for Quantum Science and Technology (Grant No. 2021ZD0301602), the National Science Association Funds (Grant No. U2230402), and the National Natural Science Foundation of China (Grant Nos. 11974040 and 12150610464).

\section*{References}
\bibliographystyle{iopart-num-long}
\bibliography{references_merged}
\end{document}